\documentclass[twocolumn]{aastex62}

\usepackage{amsmath}

\received{}
\revised{}
\accepted{}
\submitjournal{ApJ}

	\shorttitle{Energy diffusion and the deposition of energetic electron energy in flares}
	\shortauthors{Jeffrey et al.}

\begin{document}

	\title{The Role of Energy Diffusion in the Deposition of Energetic Electron Energy in Solar and Stellar Flares}

	\author[0000-0001-6583-1989]{Natasha L. S. Jeffrey}
	\affil{School of Physics \& Astronomy, University of Glasgow, G12 8QQ, Glasgow, UK}

	\author[0000-0002-8078-0902]{Eduard P. Kontar}
	\affil{School of Physics \& Astronomy, University of Glasgow, G12 8QQ, Glasgow, UK}

	\author[0000-0001-9315-7899]{Lyndsay Fletcher}
	\affil{School of Physics \& Astronomy, University of Glasgow, G12 8QQ, Glasgow, UK}

\correspondingauthor{Natasha L. S. Jeffrey}
\email{natasha.jeffrey@glasgow.ac.uk}

\begin{abstract}
During solar flares, a large fraction of the released magnetic energy is carried by energetic electrons that transfer and deposit energy in the Sun's atmosphere. 
Electron transport is often approximated by a cold thick-target model (CTTM), assuming that electron energy is much larger than the temperature of the ambient plasma, and electron energy evolution is modeled as a systematic loss. 
Using kinetic modeling of electrons, we re-evaluate the transport and deposition of flare energy. Using a full collisional warm-target model (WTM), 
we account for electron thermalization and for the properties of the ambient coronal plasma such as its number density, temperature and spatial extent. We show that the deposition of non-thermal electron energy in the lower atmosphere is highly dependent on the properties of the flaring coronal plasma. 
In general, thermalization and a reduced WTM energy loss rate leads to an increase of non-thermal energy 
transferred to the chromosphere, and the deposition of non-thermal energy at greater depths. The simulations show that energy is deposited in the lower atmosphere initially by high energy non-thermal electrons, and later by lower energy non-thermal electrons that partially or fully thermalize in the corona, over timescales of seconds, unaccounted for in previous studies. This delayed heating may act as a diagnostic of both the injected non-thermal electron distribution and the coronal plasma, vital for constraining flare energetics.
\end{abstract}

\keywords{Sun: flares -- Sun: X-rays -- Sun: atmosphere 
-- Sun: chromosphere -- Sun: corona -- stars: flare}

\section{Introduction}\label{intro}

Solar flares are a product of the Sun's magnetic energy being released and then ultimately dissipated in different layers of its vast atmosphere. The release of magnetic energy, 
initiated by magnetic reconnection in the corona \citep[e.g.,][]{1957JGR....62..509P,1958IAUS....6..123S,2000mare.book.....P}, 
is partitioned into thermal and non-thermal particle energies \citep[e.g.,][]{2012ApJ...759...71E,2015ApJ...802...53A,2017ApJ...836...17A,2016A&A...588A.116W}, and kinetic plasma motions (`turbulence'), a vital energy transfer mechanism in the process \citep[e.g., ][]{1993ApJ...418..912L,2012SSRv..173..535P,2016ApJ...827L...3V,2017PhRvL.118o5101K}. 
However, the bulk of the released energy is eventually transferred to the Sun's cool 
and dense low atmosphere (the chromosphere) causing rapid heating, 
ionization \citep[cf][]{2011SSRv..159...19F,2011SSRv..159..107H}, 
and an expansion of the lower atmospheric material - ``chromospheric evaporation'' \citep[e.g., ][]{1973NASSP.342....3S,1974SoPh...34..323H,1982ApJ...263..409A,2011SSRv..159..107H}. The heated chromosphere, a thin and complex layer, is a prime source of deposited energy information,
mainly radiating in optical and ultraviolet (UV) wavelengths \citep[e.g., ][]{1974SoPh...34..323H,2011A&A...530A..84K,2006JGRA..11110S14W}. Energy is likely transferred in a variety of different, but closely connected, ways: 
by flare-accelerated electrons, as evident from hard X-ray (HXR) observations \citep{2011SSRv..159..107H}, by thermal conduction \citep[e.g., ][]{1970SoPh...15..394C}, and complicated by various plasma waves 
e.g.  acoustic waves \citep[e.g.][]{1979ApJ...233..717V},
Alfv\'en waves \citep[e.g.][]{1982SoPh...80...99E,2008ApJ...675.1645F},
Langmuir waves \citep{1984ApJ...279..882E,1987A&A...175..255M,2009ApJ...707L..45H},
whistler waves \citep[e.g.][]{1991ApJ...374..369B,2007A&A...465..613S},
and then dissipated via turbulence, even in the lower atmosphere e.g. \citet{Jeffreyeaav2794}. 

In the flare impulsive phase, 
X-ray observations \citep{2008LRSP....5....1B,2011SSRv..159..301K} suggest that non-thermal electrons are the main source of low atmosphere heating and radiation.
Bremsstrahlung X-rays provide a relatively direct diagnostic of the properties of flare-accelerated electrons \citep[cf][]{2011SSRv..159..301K} in the corona and in the dense lower atmosphere via HXR footpoints \citep[e.g. ][]{1981ApJ...246L.155H}. Higher energy HXRs are observed 
to be produced in progressively lower regions of the chromosphere \citep{2002SoPh..210..383A,2010ApJ...717..250K} by electron-ion (mainly proton) collisions, and via electron-electron collisions above $\sim$300~keV \citep{2007ApJ...670..857K}. However,
electrons predominantly exchange energy via electron-electron collisions \citep[cf][]{2011SSRv..159..107H}.
Flare observations of `coronal thick-target' sources 
\citep[e.g.,][]{1997ApJ...487..936A,2004ApJ...603L.117V,2008ApJ...673..576X,2011ApJ...730L..22K,2012A&A...543A..53G,2013ApJ...766...75J} show that electrons with energies up to $\approx$30~keV can thermalize in the corona in high density conditions. However, more general statistical studies of large flares
\citep[e.g. ][]{2014ApJ...781...43C,2015ApJ...802...53A} 
show that the flaring corona, at least within the main phase, 
is often a highly collisional environment. 
Further, it is likely that non-collisional transport effects such as turbulent scattering by magnetic fluctuations \citep{1991ApJ...374..369B,2014ApJ...780..176K}, beam-driven Langmuir wave turbulence \citep{2009ApJ...707L..45H}, 
electron re-acceleration \citep{2009A&A...508..993B} 
and/or beam-driven return current \citep{1977ApJ...218..306K,1980ApJ...235.1055E,2006ApJ...651..553Z,2017ApJ...851...78A} are also operating during flares, complicating the overall transport.

For the last fifty years, the properties of non-thermal electrons (their transport, deposition and the heating of the lower atmosphere), 
are often determined using the `cold-thick-target' collisional transport 
model (hereafter CTTM) \cite[e.g., ][]{1971SoPh...18..489B,1972SvA....16..273S,1978ApJ...224..241E}. The CTTM assumes that the energy $E$ of non-thermal electrons is much larger 
than the ambient plasma temperature $T$, and hence `cold' (i.e., $T \ll E$). Although this assumption is valid for high-energy electrons 
that reach the cool layers of the flaring chromosphere, 
decades of observational evidence with e.g., \textit{Yohkoh} \citep{1991SoPh..136...37T} 
and the \textit{Reuven Ramaty High Energy Solar Spectroscopic Imager} (\textit{RHESSI}; \citet{2002SoPh..210....3L}) 
show high coronal temperatures of 10-30~MK during flares. 
However, the lasting appeal of the CTTM is its simple analytic form, 
that can be readily applied to X-ray data, but its use leads to the well-known `low-energy cut-off' problem, 
whereby the power associated with non-thermal electrons 
cannot be constrained from X-ray spectroscopy. Firstly, \citet{2014ApJ...787...86J}, building upon \citet{2003ApJ...595L.119E} and \citet{2005A&A...438.1107G}, studied electron transport using a full collisional model including finite temperature effects, diffusion and pitch-angle scattering, and showed the importance of including the properties of the coronal plasma (its finite temperature, density and extent). 
Critically, the inclusion of both thermalization and spatial diffusion 
led to the ``warm-target model'' (hereafter WTM; derived by \citet{2015ApJ...809...35K})
that can resolve the problems associated with determining the low-energy cut-off in the CTTM, finally allowing the power of flare-accelerated electrons to be constrained \citep{2019ApJ...871..225K} from X-ray data. In a WTM, the properties and energy content of non-thermal electrons are \textit{constrained by determining the plasma properties of the flaring corona}.

Here, using full collisional kinetic modelling, we re-investigate flare-accelerated electron energy deposition. As expected, we show that the coronal plasma properties (e.g. temperature, number density and spatial extent) determine how non-thermal electron power is deposited in the chromosphere. Ultimately, we show for a given non-thermal electron distribution, a greater proportion of the non-thermal electron power can be deposited in the lower atmosphere than predicted in the CTTM.

	\begin{figure*}[t]
	\centering
	\includegraphics[width=0.80\linewidth]{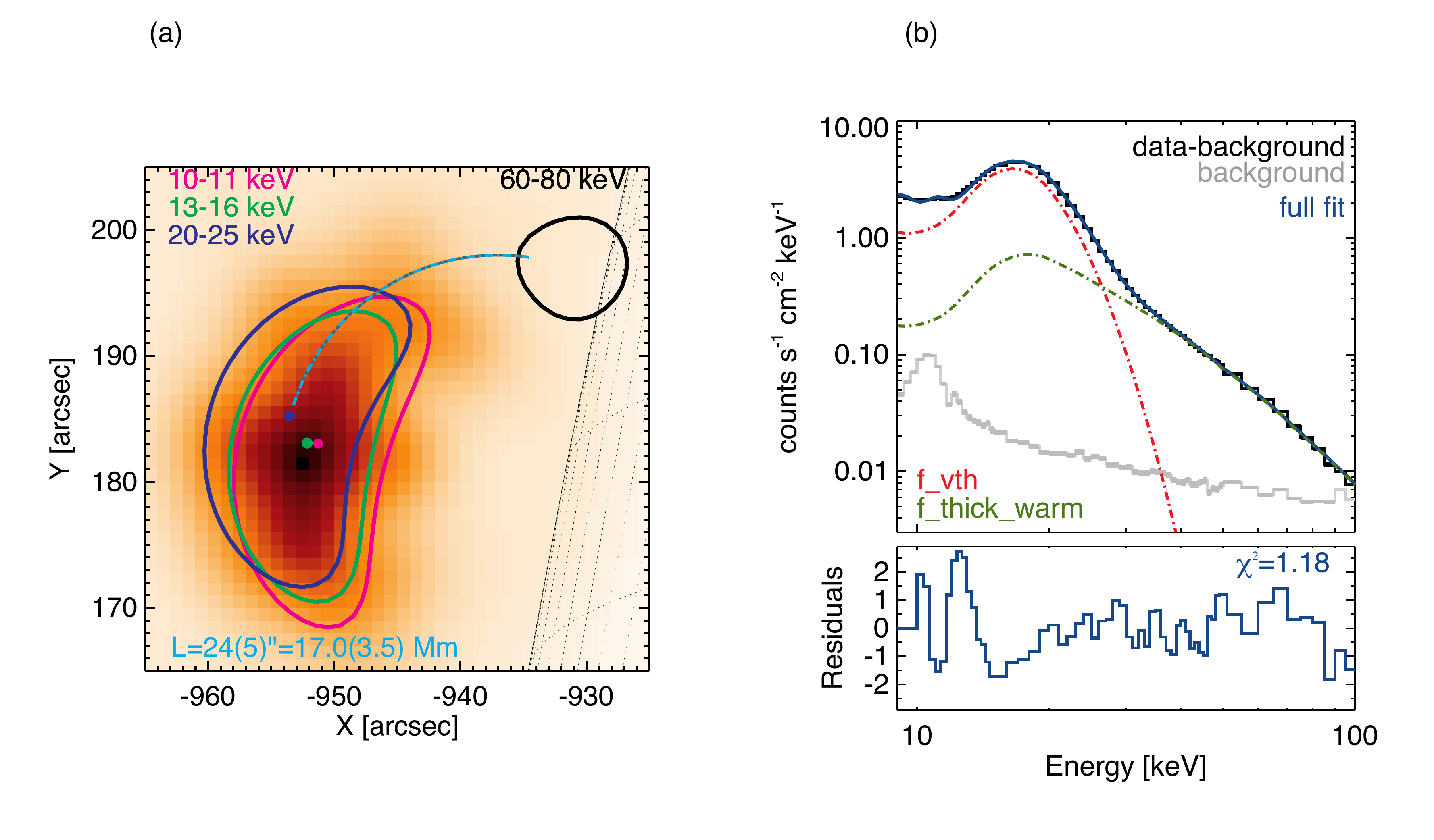}
	\caption{A {\em RHESSI} X-ray observation of a flare (SOL2013-05-13T02:12). (a)  We observe lower energy X-rays from a hot thermal source in the corona and higher energy X-rays from accelerated electrons reaching the cooler and denser chromosphere. (b) The total spatially integrated X-ray count spectrum fitted with a thermal component (red) and a WTM component (green) that accounts for both high energy electrons (power law) and low energy thermalized electrons. From combined X-ray spectroscopy and imaging, we determine that the coronal source is hot and dense with $\langle T\rangle\approx29$~MK and $\langle n\rangle\approx9\times10^{10}$~cm$^{-3}$. Here, the distance between the X-ray coronal and footpoint sources is $L\approx24 \arcsec$. The determination of these coronal plasma properties is vital for constraining electron transport and deposition. Figures taken from \citet{2019ApJ...871..225K}.}
	\label{fig0}
	\end{figure*}

	\begin{figure}[t]
	\centering
	\includegraphics[width=0.95\linewidth]{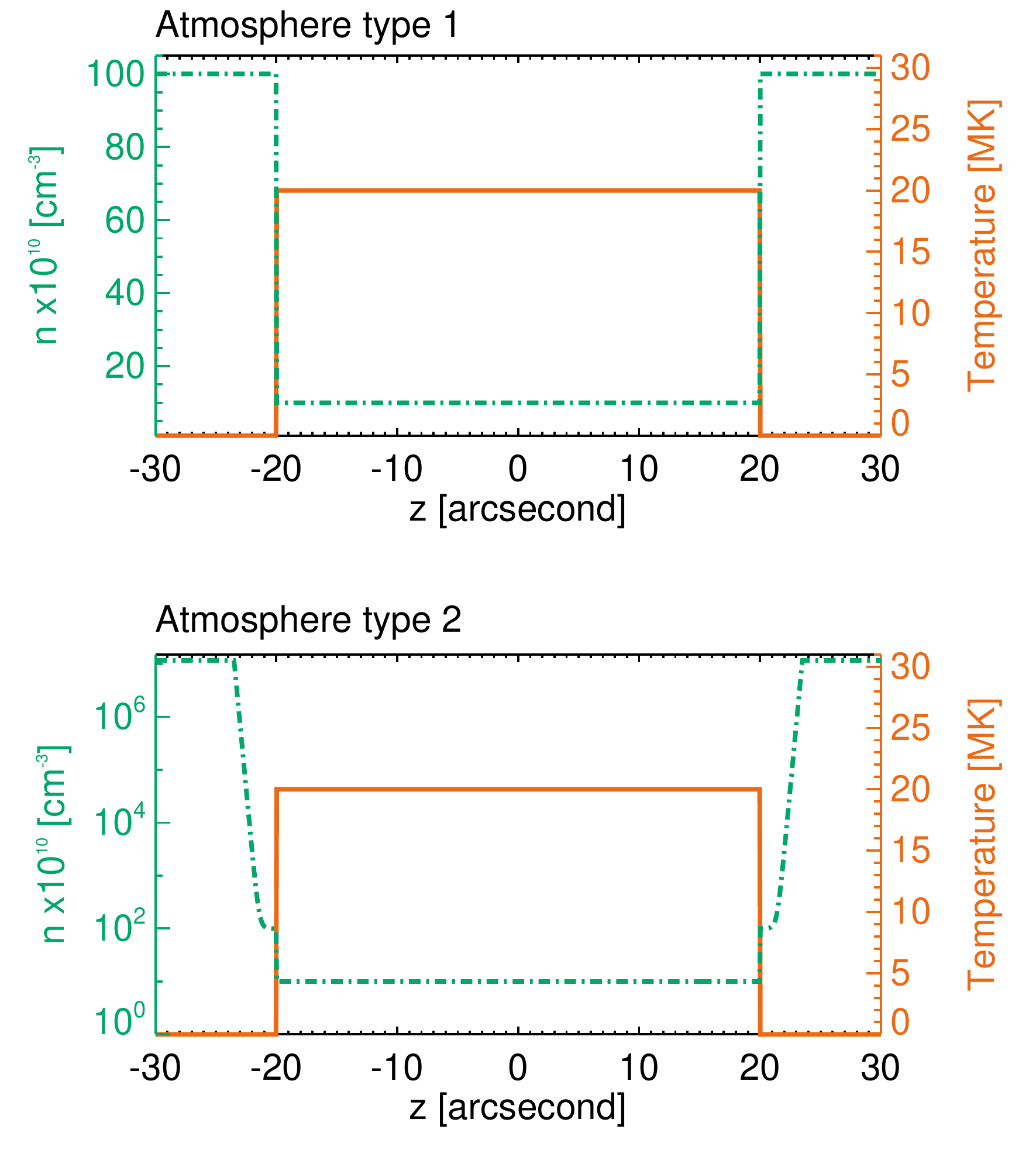}
	\caption{The different atmospheric types used in the simulations. \textit{Top:} Atmosphere 1. - a constant coronal temperature (we use either 10, 20, 30 or 100~MK) and number density (either $3\times10^{10}$, $7\times10^{10}$ or $1\times10^{11}$~cm$^{-3}$) and `cold-target' chromosphere with $n=10^{12}$~cm$^{-3}$. \textit{Bottom:} Atmosphere 2. - as Atmosphere 1. with a constant coronal temperature and number density, but with an exponentially increasing chromospheric density (see Section \ref{plasma}), with a scale height of $130$~km and photospheric density of $n_{\rm photo}=1.16\times10^{17}$~cm$^{-3}$. The coronal plasma has a half-loop length of either $L=20\arcsec$ or $L=30\arcsec$.}
	\label{fig1}
	\end{figure}

\begin{figure*}[t]
	\centering
	\includegraphics[width=0.6\linewidth]{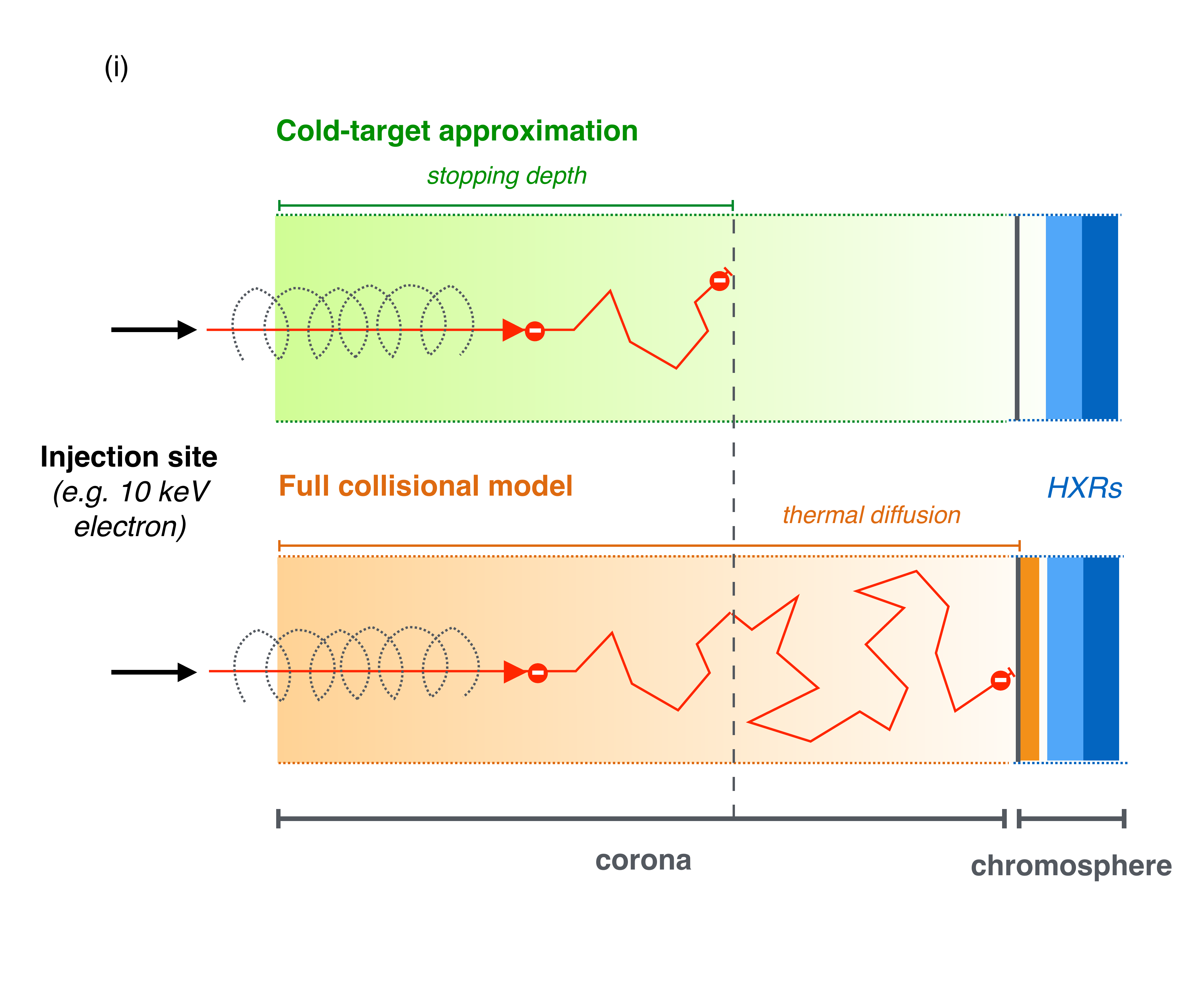}
	\includegraphics[width=0.9\linewidth]{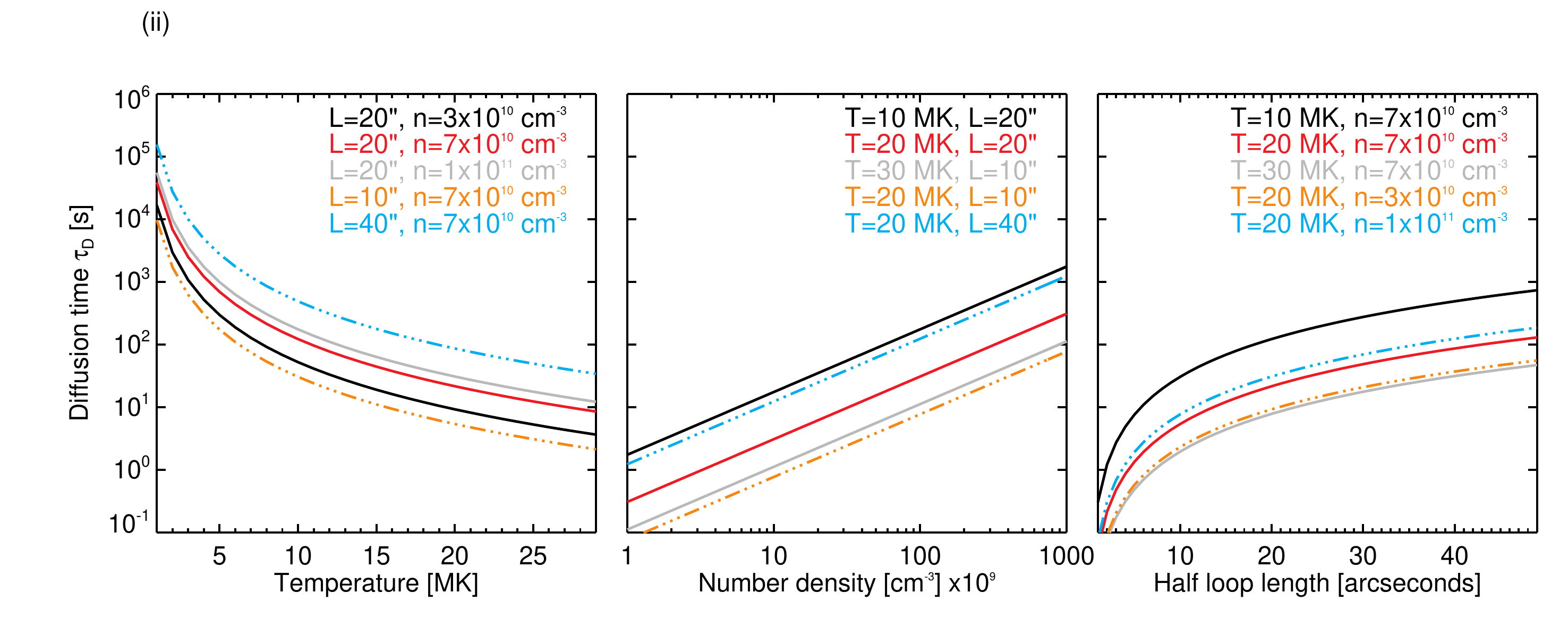}
		\caption{(i) A cartoon showing the difference between low-energy electron transport in a CTTM and full collisional model (WTM). In a CTTM, low-energy electrons might never reach the chromosphere, if they are collisionally stopped in the corona, but in a WTM, such electrons are thermalized and can still transport a fraction of their non-thermal energy from the hot corona to the cool chromosphere via thermal diffusion (orange layer in the chromosphere).
		(ii) Comparison of the time it takes a thermalized electron to spatially diffuse from a hot coronal source to a cooler and denser low atmosphere, using different coronal plasma conditions of number density $n$, temperature $T$, and the distance between the coronal source and the chromosphere $L$. For solar flare conditions, i.e. $T\approx10-30$~MK, $n\approx10^{10}-10^{11}$~cm$^{-3}$ and $L\approx20-30\arcsec$, $\tau_{D}$ ranges between $1-100$~s. Such timescales are important in the WTM where non-thermal electrons partially or fully thermalize during their transport in the corona and hence, transfer energy to the chromosphere on timescales of seconds.}
	\label{fig2}
	\end{figure*}

\section{Electron transport and deposition in hot collisional plasma}\label{WTM}

To determine how the energy of flare-accelerated electrons is transported and deposited in a hot collisional plasma (i.e. in a full collisional WTM), 
we use the kinetic electron transport simulation first discussed in \citet{2014ApJ...787...86J} and \citet{2015ApJ...809...35K}. 
We model the evolution of an electron flux 
$F(z,E,\mu)$~[electron erg$^{-1}$ s$^{-1}$ cm$^{-2}$] in space $z$~[cm], energy $E$~[erg], and pitch-angle $\mu$ to a guiding magnetic field, 
using the Fokker-Planck equation of the form \citep[e.g.,][]{1981phki.book.....L,1986CoPhR...4..183K}:

\begin{equation}\label{eq: fp_e}
\begin{split}
\mu \frac{\partial F}{\partial z} &= \Gamma m_{e}^2 \left ( \frac{\partial}{\partial E} \left[ G (u[E] ) \frac{\partial F}{\partial E} + \frac{G (u[E] )}{E} \left ( \frac{E}{k_B T}-1 \right )F \right]\right) \\
&+ \Gamma m_{e}^2\left(\frac{1}{8E^2} \frac{\partial}{\partial \mu} \left [ (1-\mu^{2}) \biggl ( {\rm erf} (u[E] ) - G (u[E] ) \biggr ) \frac{\partial F}{\partial \mu} \right ] \right ) \\
& + S(E,z,\mu),\,\,\,\,
\end{split}
\end{equation}
\\
where $\Gamma=4\pi e^{4} \ln\Lambda \, n /m_{e}^{2}=2Kn/m_{e}^{2}$, 
and $e$ [esu] is the electron charge, 
$n$ is the plasma number density [cm$^{-3}$] (a hydrogen plasma is assumed), $m_{e}$ is the electron rest mass [g], and $\ln\Lambda$ is the Coulomb logarithm. The variable $u(E)=\sqrt{E/k_B T}$, where $k_{B}$~[erg K$^{-1}$] is the Boltzmann constant and $T$~[K] is the background plasma temperature. The functions ${\rm erf}(u)$ (the error function) and $G(u)$ are given by,

\begin{equation}\label{eq:gcha}
{\rm erf}(u)\equiv (2/\sqrt{\pi})\int\limits_{0}^{u}\exp(-t^2) \, dt
\end{equation}
and
\begin{equation}
G(u)=\frac{{\rm erf}(u)-u \, {\rm erf}^{'}(u)}{2u^{2}} \,\,\, .
\end{equation}

Equation (\ref{eq: fp_e}) is a time-independent equation useful for studying solar flares where the electron transport time from the corona to the lower atmosphere is usually shorter than the observational time 
(i.e. most X-ray observations have integration times of tens of seconds to minutes). 

Here, Equation (\ref{eq: fp_e}) models electron-electron collisions only,
the dominant electron energy loss mechanism in the flaring plasma\footnote{We note that electron-proton interactions are important for collisional pitch-angle scattering, but here we only model electron-electron interactions. Equation (\ref{eq: fp_e}) can be generalized to model any particle-particle collisions.}.
$S(E,z,\mu)$ plays the role of the electron flux source function 
and the properties of the injected electron distribution 
are discussed in Section \ref{inject}.

Following \citet{2014ApJ...787...86J}, and re-writting Equation (\ref{eq: fp_e}) 
as a Kolmogorov forward equation \citep{Kolmogorov1931}, 
Equation (\ref{eq: fp_e}) can be converted to a set of time-independent 
stochastic differential equations (SDEs) 
\citep[e.g., ][]{1986ApOpt..25.3145G,2017SSRv..212..151S} 
that describe the evolution of $z$, $E$, and $\mu$ in It$\hat{\text{o}}$ calculus:

\begin{equation}\label{eq:sto_x}
z_{j+1}=z_{j}+\mu_{j} \, \Delta s \,\,\, ;
\end{equation}
\begin{equation}\label{eq:sto_E}
\begin{split}
E_{j+1} & =E_{j}-\frac{\Gamma m_{e}^{2}}{2E_{j}} \, \bigg ( {\rm erf}(u_{j})-2u_{j}{\rm erf^{\prime}}(u_{j}) \bigg) \, \Delta s\\
& +\sqrt{2 \, \Gamma m_{e}^{2} \, G(u_{j}) \, \Delta s} \,\, W_E \,\,\, ; \,\,\,
\end{split}
\end{equation}
\begin{equation}\label{eq:sto_mu}
\begin{split}
\mu_{j+1}&=\mu_{j}-\frac{\Gamma m_{e}^{2} \biggl ( {\rm erf}(u_{j})-G(u_{j}) \biggr ) } {4 E_{j}^{2}} \,\, \mu_{j} \, \Delta s \\
&+\sqrt{\frac{ (1-\mu_{j}^{2}) \, \Gamma m_{e}^{2} \, \biggl ( {\rm erf}(u_{j})-G(u_{j}) \biggr ) } {4 E_{j}^{2}} \, \Delta s} \, \, W_\mu \,\,\, .
\end{split}
\end{equation}
$\Delta s$~[cm] is the step size along the particle path, 
and $W_\mu$, $W_E$ are random numbers drawn from Gaussian 
distributions with zero mean and a unit variance representing the corresponding Wiener processes \citep[e.g. ][]{1986ApOpt..25.3145G}. A simulation step size of $\Delta s=10^{5}$~cm is used in all simulations, and $E$, $\mu$ and $z$ are updated at each step $j$. A step size of $\Delta s=10^{5}$~cm is approximately two orders of magnitude smaller than the thermal collisional length in a dense ($n=10^{11}$~cm$^{-3}$) plasma with $T\ge 10$~MK (or the collisional length of an electron with an energy of $1$~keV or greater, in a cold plasma). The simulation ends when all `electrons' have left the warm-target corona, and reached the cool `chromosphere' (where a CTTM approximation is valid for all studied energies). Using a time-independent equation with a constant source of injection produces output variables with units of [electron s$^{-1}$ per output variable] i.e., $E$, $z$ or $\mu$, and the final results are reconstructed by summing over all outputs at each step $j$.
The derivation of Equation (\ref{eq: fp_e}) and the detailed 
description of the simulations can be found in \citet{2014ApJ...787...86J}. 

Equation (\ref{eq: fp_e}) (and Equations (\ref{eq:sto_E}) and (\ref{eq:sto_mu})) diverge as $E\rightarrow0$, and as discussed in \citet{2014ApJ...787...86J}, 
the deterministic equation $E_{j+1}=\left[E_{j}^{3/2} + \frac{3\Gamma m_{e}^{2}}{2\sqrt{\pi k_{B}T}}\Delta s \right]$ must be used 
for low energies where $E_{j}\le E_{\rm low}$ using $E_{\rm low}=\left[\frac{3\Gamma m_{e}^{2}}{2\sqrt{\pi k_{B} T} \Delta s}\right]^{2/3}$ -- see \citet{2014ApJ...787...86J}, following \citet{2009JCoPh.228.1391L}. For such low energy thermal electrons, $\mu_{j+1}$ can be drawn from an isotropic distribution $\mu \in [-1,+1]$.

\subsection{The deposition of non-thermal electron power}\label{deposition}

Electron energy deposited into the ambient plasma can be determined by considering
\begin{equation}\label{eq:d1}
\Delta E_{j+1} (z)=E_{j} (z)-E_{j+1} (z),
\end{equation}
where $E_{j}$ and $E_{j+1}$ are the electron energies before and after each simulation step
respectively. Using $\Delta E_{j+1} (z)$, a new ambient background temperature at that location can also be determined. However, we do not examine changes in background plasma temperature, and the background temperature remains constant in all simulations.

Although derived from a time-independent equation, we note that Equations (\ref{eq:sto_x}-\ref{eq:sto_mu}) are related to a time step $\Delta t$ by $\Delta t=\Delta s/v$, where $v=\sqrt{2E/m_{e}}$ is the velocity of the electron, and the simulations can also be used for a \textit{time-dependent} analysis. In all simulations, the total time it takes for an electron to deposit all of its energy can be approximated using 
\begin{equation}\label{eq: depo_tim}
t=\sum_{j}\frac{|z_{j+1}-z_{j}|}{\sqrt{2E_{j}/m_{e}}}.
\end{equation} 
In cases where $\mu_{j}\approx0$, we use $|z_{j+1}-z_{j}|\approx \Delta s$. Also in the CTTM, this time can be estimated analytically using $t=\frac{E_{0}^{2}}{2Knv}$ where $E_{0}$ is the injected energy of the electron. 

During each simulation step $j$, in order to determine the non-thermal electron 
power $P_{j+1}(z)$ at each spatial location $z$,
we weight the output of [electron s$^{-1}$ cm$^{-1}$] in each $z$ bin by the total $\Delta E_{j+1}$ deposited in that bin, giving $P_{j+1}(z)$ [erg s$^{-1}$ cm$^{-1}$]. 
Summing over all saved $j$ gives the total non-thermal electron power $P(z)$ deposited at each spatial location $z$, equivalent to
\begin{equation}\label{eq:d1}
P(z)=\int_{0}^{\infty}E F(E,z) dE.
\end{equation}
Further, summing over all $z$ gives the total spatially integrated non-thermal electron power [erg s$^{-1}$], which can be compared with the injected non-thermal electron power $P$ [erg s$^{-1}$] input into the simulation. For an injected non-thermal electron power law distribution \citep[cf. ][]{2011SSRv..159..107H}, this can be written as
\begin{equation}
P=\int_{E_{c}}^{\infty}E_{0}F(E_{0})dE_{0}=\dot{N_{0}} E_{c}\frac{(\delta-1)}{(\delta-2)},
\end{equation}
for the injected energies $E_{0}$, an acceleration rate $\dot{N_{0}}$ [electron s$^{-1}$], a low energy cutoff of $E_{c}$ (the lowest energy in the non-thermal electron distribution), and the power law spectral index $\delta$.

\subsection{Flare plasma parameters}\label{plasma}

We model the flaring atmosphere using a {\it hot corona--cold chromosphere} type model (see Figures \ref{fig0} and \ref{fig1}). This atmosphere is a simple but reasonable description of most flaring atmospheres. 
Moreover, a more realistic atmosphere is not required since we only want 
to {\it compare} the results of the CTTM and WTM. This type of atmosphere 
also ensures that a time-independent stationary solution is reached \citep[e.g., see ][for details]{2015ApJ...809...35K}. Unlike the CTTM, electrons are no longer lost energetically, but accumulate in the corona as they thermalize. This pile-up of thermalized electrons in the corona is balanced by the spatial diffusion of electrons from the hot corona 
into the cool chromosphere, which can be still considered a cold-target.

We perform simulation runs for two different ``hot corona -- cool chromosphere'' model atmospheres (see Figure \ref{fig1}), including one model that includes a chromosphere with an exponential density profile \citep[e.g. ][]{1981ApJS...45..635V,2012ApJ...752....4B}.

The development of the WTM has shown that the plasma parameters
(the coronal temperature $T$, the coronal number density $n$,
and the coronal plasma extent $L$, where the temperature is high enough
to be visible in X-rays) are crucial for determining and constraining 
the properties of flare-accelerated electrons \citep{2019ApJ...871..225K}. Here, we show how the plasma properties play a key role 
in the transfer and the deposition of non-thermal electron power.
We test how the energy of non-thermal electrons is transferred 
and deposited in a range of different coronal plasma conditions. 
In the corona, we use different number densities ranging 
from $n=3\times10^{10}$~cm$^{-3}$ to $n=1\times10^{11}$~cm$^{-3}$,
and plasma extents (half-loop lengths $L$) of either 20\arcsec or 30\arcsec\,(see Figure \ref{fig0}) between the hot corona and cooler chromosphere, 
leading to column depths of $10^{19}-10^{20}$~cm$^{-2}$. 
In the WTM cases, coronal temperatures range from 10~MK to 30~MK 
(see Figure \ref{fig0}) for solar/M-dwarf cases and up 
to 100~MK for comparison with certain extreme 
stellar cases \citep[e.g. see Figure 3 in ][]{2008ApJ...672..659A}.
In atmosphere type 1, in the cool `chromosphere-type' region, 
the number density rises to $n=1\times10^{12}$~cm$^{-3}$ 
and the temperature falls to $T\sim0$~MK, i.e., it is approximated as a CTTM. In the more realistic atmosphere type 2, the density at the boundary 
of the cool ``chromosphere-type'' region is set at $n=1\times10^{12}$~cm$^{-3}$, 
but this rises quickly to photospheric densities 
of $n=\sim10^{17}$~cm$^{-3}$ over $\sim3\arcsec$ using
\begin{equation}\label{atmos2}
n=10^{12}\,{\rm[cm^{-3}]}+n_{\rm photo}\exp{\left(-\frac{|z|+23\arcsec.5}{h_{0}}\right)},
\end{equation}
where $n_{\rm photo}=1.16\times10^{17}$~cm$^{-3}$ 
is the photospheric density at the optical depth of $\approx1$, 
and here $z$ is measured in arcseconds. 
The scale height $h_{0}$ of the density profile is set at $0\arcsec.18\sim130$~km \citep[e.g. following the simulations of][]{2012ApJ...752....4B}. 

In most solar flare coronal conditions, $\ln\Lambda\approx20$, 
but we can calculate $\ln\Lambda$  using \citep[e.g. ][]{Somov2007},
\begin{equation}\label{eq: co_log}
\ln\Lambda=\ln\left[\frac{3}{2 e^{3}}\left(\frac{k_{B}^{3}T^{3}}{\pi n}\right)^{1/2}\right].
\end{equation}
 
In the CTTM simulations, we choose $T=T_{\rm corona}$, 
where $T_{\rm corona}$ is the background corona temperature used 
in the WTM simulations. 
In the lower `cold-target' atmosphere, 
we choose $T=0.01$~MK for the calculation of $\ln\Lambda$.

\subsection{The injected electron distribution}\label{inject}

The source function $S(E,z,\mu)$ is made up of three separate distributions:

\begin{enumerate}
\item {\it Injected energy spectrum} -- we input either a mono-energetic distribution or a power law distribution of the form $\sim E_{0}^{-\delta}$. Electron distributions with approximate power law forms are routinely observed via X-ray observations. In the mono-energetic cases, we input electrons with energies of either, the coronal thermal energy, 10~keV, 20~keV, 30~keV, 40~keV, 50~keV or 100~keV. In these cases, we compare the outputs using $\dot{N_{0}}=1$ s$^{-1}$. For the power law case, parameters of $\delta=4$ or $\delta=7$, a low energy cutoff of $E_{c}=10$~keV or $E_{c}=20$~keV and a high energy cutoff of $E_{H}=50$~keV are used\footnote{Electrons with energies above $\sim50$~keV will approximate the CTTM solution using the noted plasma parameters, and low-energy electrons carry the bulk of the power due to steeply decreasing power laws.}. In power law cases, the electron injection rate $\dot{N_{0}}$ is set at a value that gives the total injected electron power $P=4.8\times10^{27}$~erg s$^{-1}$.

\item {\it Injected pitch-angle distribution} -- we input a beamed distribution (with half moving in one direction and half moving in the opposite direction, i.e $\mu=+1$ or $\mu=-1$). We also run simulations using a completely isotropic distribution ($\mu\in[-1,+1]$), but for brevity the results are not shown here. In general, the injected pitch-angle distribution is not well-constrained by current solar flare observations \citep[e.g. ][]{2011SSRv..159..301K,2017A&A...606A...2C}. Collisional (electron-electron only) pitch-angle scattering is always modeled in the simulations. 
    Further, it is very likely that other non-collisional (and shorter timescale) turbulent scattering mechanisms are also presence in the flaring atmosphere \citep{2014ApJ...780..176K}. This will also change how the electron energy is deposited spatially and temporally, and the subject of ongoing work. 
\item {\it Injected spatial distribution} -- we input the electrons as a Gaussian distribution centred at the loop top apex ($z=0\arcsec$), 
    with a standard deviation of 1\arcsec. 
    It is possible that electrons are accelerated to varying levels (dependent on the plasma conditions) at multiple points 
    along a twisted loop \citep[e.g. ][]{2012SoPh..277..299G,2014A&A...561A..72G}, but again simulating all possible cases is beyond the scope of the paper and it is not required for a CTTM and WTM comparison.
\end{enumerate}

\subsection{Timescales for the deposition of non-thermal electron power}\label{timescale}
 
Unlike the CTTM case, in a WTM, electrons `stopped' in the coronal plasma thermalize and then diffuse through the coronal region 
in a random walk continuously exchanging energy with the background population. Ultimately, this means that non-thermal electrons fully thermalized in the hot corona  
still transfer some fraction of their injected energy to the cool lower atmosphere (see the cartoon in Figure \ref{fig2} (i)). 

Moreover, the time is takes for an injected \textit{thermal} electron, 
which is dominated by diffusion, to leave the corona 
and deposit its energy in a cool low atmosphere (cold-target) 
can be estimated analytically using

\begin{equation}\label{diff}
\tau_{D}=\sqrt{\frac{8m_{e}}{\pi}}\;\frac{KnL^{2}}{3} (k_{B}T)^{-5/2}
\end{equation}

where $K=2\pi e^{4} \ln{\Lambda}$. This diffusion time ($\tau_{D}$; Equation \ref{diff}) is comparable to the Spitzer thermal conduction time \citep{1962pfig.book.....S}. In Figure \ref{fig2} (ii), 
we calculate $\tau_{D}$ for a range of different coronal 
parameters: $T$, $n$ and $L$. $\tau_{D}$ increases with increasing $n$ 
and $L$, and decreases with increasing $T$ 
(thermalized electrons in hotter plasma have a higher thermal energy 
and hence reach the chromosphere quicker). 
For the range of plasma parameters used in the simulations, 
$\tau_{D}$ ranges between $\sim1-100$~s. 
Although, the thermal electron diffusion time $\tau_{D}$ 
can be calculated analytically, it is not trivial to determine the deposition 
timescales for non-thermal electrons injected with $E>k_{B}T$ into different coronal plasma conditions. 
Therefore, in each simulation run, we determine the time it takes for non-thermal electrons of different injected energies to deposit their energy in the lower atmosphere using Equation ({\ref{eq: depo_tim}).

In these simulations, we stress that we do not consider the energy transferred from the hot corona to the cool lower atmosphere from the background thermal plasma by thermal conduction, which will play a varying role in different flares and at different stages as the flare progresses. Here, we only consider the energy transferred by non-thermal electrons, and in particular, the `extra' component of energy that comes from partially or fully thermalized non-thermal electrons now being able to reach the lower atmosphere from diffusion.

In the simulations, we make several simplifying assumptions, such as: (1) use of single temperature and number density in the corona, (2) a previously heated corona and the use of a coronal `thermal bath' approximation, i.e. there are no significant changes in the background energy content due to the transport and deposition of non-thermal electron energy, (3) the use of a `step function' type atmosphere. Here, the assumptions are valid for a CTTM and WTM comparison, as stated. Further, the thermal diffusion of electrons is a fundamental transport mechanism always present in flares (to a varying degree depending on the coronal environment), that is usually overlooked, even in Fokker-Planck type simulations where a full collisional model is used. Hence, more complex simulations are not required here, and they would hinder our comparison of energy transport in the CTTM and WTM, which is the main aim of this study.

\section{Simulation results}\label{results}

\subsection{Mono-energetic energy inputs: spatial distribution of deposited power}\label{monos}

Firstly, we perform simulations where we input different mono-energetic electron distributions into different plasma environments, so that the results of the WTM 
and CTTM can be easily compared for electrons of different energies in a range of different coronal conditions. 
Here, in each simulation run, the accelerated electron rate is $\dot{N_{0}}=1$ s$^{-1}$, for each input. We perform four different sets of simulations 
labelled (a)-(d), using atmosphere 1 (see Figure \ref{fig1}):

\textit{Sets (a) and (b): different coronal densities-} In sets (a) and (b), we input mono-energetic electrons with energies of either 10~keV, 20~keV, 30~keV, 40~keV, 50~keV or 100~keV. 
The injected electrons are initially beamed in $\mu$, and spread in $z$ as a Gaussian with a $1''$ standard deviation centered at the loop apex. 
They are injected into atmosphere type 1 (see Figure \ref{fig1}), 
with either a coronal density of (a) $n=10^{11}$~cm$^{-3}$ (Figure \ref{fig3}), or (b) $n=3\times10^{10}$~cm$^{-3}$ (Figure \ref{fig4}), a half loop length of $L=20\arcsec$, and for all the WTM cases, a coronal temperature of $T=20$~MK (giving $E/k_{B}T=5.8, 11.6, 17.4, 23.2, 29.0, 57.9$). All the WTM and CTTM results are shown in Figure \ref{fig3} and Figure \ref{fig4} (showing the spatial distribution of non-thermal power deposition in units of [keV s$^{-1}$ arcsecond$^{-1}$] and the ratio of the WTM to CTTM result for each run). Tables \ref{tb1} and \ref{tb2} show the percentage of non-thermal power deposited in the corona or low atmosphere 
in both WTM and CTTM cases. From the results of (a) and (b) we find:
\begin{enumerate}
\item In (a) and (b), the spatial distribution of deposition in the WTM is different from the CTTM for all energies up to and including $\approx50$~keV, and this difference increases for {\it larger coronal densities} and {\it smaller injected electron energies} (see Figure \ref{fig3} and Figure \ref{fig4}).
\item In (a) and (b), electrons can move further and deposit more energy at greater depths in both the corona and low atmosphere in the WTM. This is most obvious for low energy electrons ($<30$~keV) in a high density corona. Such electrons are collisionally stopped in the corona. The CTTM predicts that they deposit all of their energy in the corona. In the WTM, electrons thermalize (or tend towards a Maxwellian) and eventually deposit some fraction of their original energy content in the low atmosphere (see Table \ref{tb1} and Table \ref{tb2} for a comparison of the CTTM and WTM percentages). For example, in (a), an injected 20~keV electron population now deposits $\sim20\%$ of its available non-thermal power in the lower atmosphere, compared to $0\%$ in a CTTM.
\item The ratio of WTM to CTTM deposition is an informative parameter that shows, in a single given location, up to 100 times more non-thermal electron power can be deposited (in both the corona and low atmosphere), than predicted by the CTTM, for a given injection of non-thermal electrons (see Figure \ref{fig3} and Figure \ref{fig4}).
\end{enumerate}

\textit{Set (c): different coronal temperatures-} In set (c), we compare WTM and CTTM deposition in different coronal plasma temperatures of $T=10,20,30,100$~MK. We inject initially beamed, mono-energetic electrons with an energy of 30~keV only (giving $E/k_{B}T=34.8,17.4,11.6,3.5$), into atmosphere type 1, using $n=7\times10^{10}$~cm$^{-3}$ and $L=20\arcsec$. The results are shown in Figure \ref{fig5} (left) and Table \ref{tb3}. From the results of set (c), we find that:
\begin{enumerate}
\item For 30~keV electrons, the higher the temperature of the coronal plasma, the greater the fraction of non-thermal electron power transferred to the lower atmosphere, with less deposition in the corona (see Table \ref{tb3}). 
    Electrons tend to a Maxwellian, and in higher coronal temperatures, thermalize at higher energies carrying a higher fraction of their power into the lower atmosphere. This dependence on coronal temperature is completely ignored in the CTTM.
\item Using atmosphere 1, at higher coronal temperatures, more power is deposited at greater depths in the lower atmosphere, and the ratio of WTM to CTTM deposition shows that at certain $z$ in the lower atmosphere, more than three orders of magnitude more power can be deposited (see Figure \ref{fig5} (left)). In the WTM, the decreased energy loss rate in the corona means that electrons carry more energy when they reach the chromospheric boundary and hence, they can travel deeper into the lower atmosphere.
\item{It is possible that in high temperature plasma, if a fraction of the injected electrons have energies lower than $k_{B}T$, then the electrons will thermalize in the corona \textit{gaining energy} and hence, deposit a higher fraction of energy in the lower atmosphere than suggested by their initial injected energy.}
\end{enumerate}

\textit{Set (d): different coronal loop lengths-} In set (d), we compare WTM and CTTM deposition in different loop lengths of $L=20\arcsec$ and $L=30\arcsec$. We inject beamed, 
mono-energetic electrons of 30~keV only into atmosphere type 1, using $n=7\times10^{10}$~cm$^{-3}$ and $T=20$~MK. The results are shown in Figure \ref{fig5} (right) and Table \ref{tb3}. From the results of set (d), we find that:
\begin{enumerate}
\item As expected, irrespective of the coronal loop length, more power is transferred to the lower atmosphere in the WTM compared to the CTTM. The larger the loop length (greater column depth), the greater time electrons spend 
    in the corona and hence, more electrons tend towards Maxwellian or fully thermalize in the corona.
\item Although, more power reaches the lower atmosphere in smaller loop lengths, the difference between the power transferred in the WTM and CTTM is greater for larger loop lengths. For example, for $L=30\arcsec$, only 0.7\% of the total non-thermal power is transferred in the CTTM but in the WTM this rises to 28\%. Further, the ratio of WTM/CTTM deposition at the chromospheric boundary is $\sim100$.
\end{enumerate}

	\begin{figure*}[thpb]
	\centering
	\includegraphics[width=1.00\linewidth]{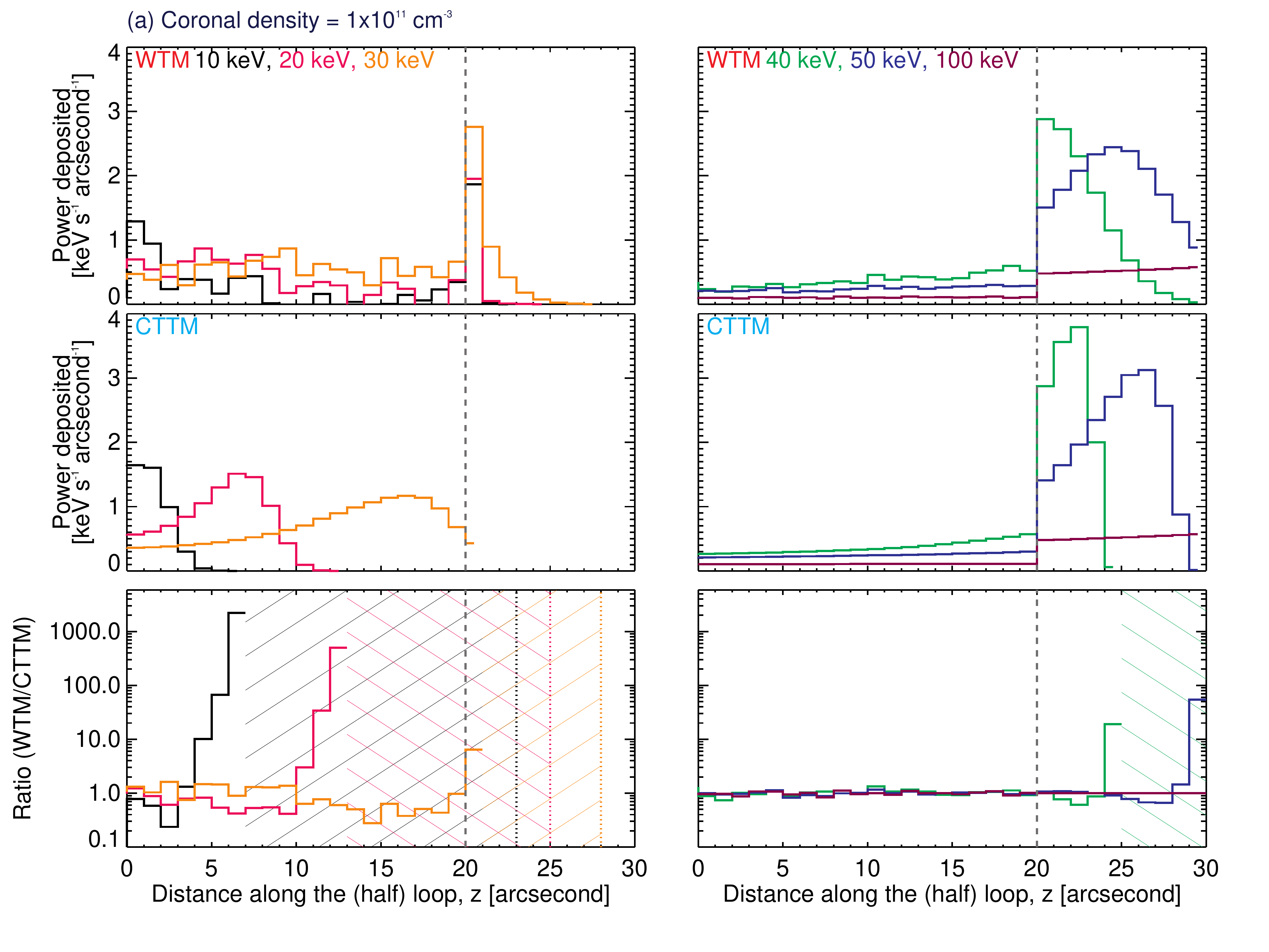}
	\caption{{\it High density flaring corona}. The results for the (initially) mono-energetic simulation set (a) showing the spatial distribution of the electron power deposition in a WTM (top) and CTTM (middle), for different injected electron energies (10, 20, 30, 40, 50, 100 keV). The ratio of WTM result to CTTM result is shown in the bottom panel. The shaded regions in the bottom panels show regions where the power is deposited at greater depths in the WTM compared to the CTTM (and hence ratio$\rightarrow\infty$). Dashed grey line: corona-chromosphere boundary. Simulation set (a) uses a coronal temperature of $T=20$~MK, coronal loop length of $L=20\arcsec$, and a beamed injection.}
	\vspace{-10pt}
	\label{fig3}
	\end{figure*}
	\begin{table*}[thpb!]{}
	\begin{center}
	\footnotesize
	\begin{tabular}{|c|cc|cc|}
	\hline
	Set (a)  & \multicolumn{2}{c|}{CTTM} & \multicolumn{2}{c|}{WTM (T=20~MK)} \\ \hline
	\begin{tabular}{l}Energy {[keV] (E/T)}\end{tabular} & \begin{tabular}{l}Chromosphere ($\%$)\end{tabular} & \begin{tabular}{l}Corona ($\%$) \end{tabular} & \begin{tabular}{l}Chromosphere ($\%$) \end{tabular} & \begin{tabular}{l}Corona  ($\%$)\end{tabular} \\ \hline
	10 (5.8) & 0 & 100 & 61.1 & 38.9 \\
	20 (11.6) & 0 & 100 & 21.3 & 78.7 \\
	30 (17.4) & 5.3 & 94.7 & 31.7 & 68.3 \\
	40 (23.2) & 64.2 & 35.8 & 63.7 & 36.3 \\
	50 (29.0) & 80.7 & 19.3 & 79.6 & 20.4 \\
	100 (57.9) & \textit{88.7} & \textit{11.3} & \textit{88.7} & \textit{11.3}\\
	\hline
	\end{tabular}
	\caption{The percentage of available non-thermal electron power deposited in the corona and chromosphere for set (a) shown in Figure \ref{fig3}. The 100~keV values are shown in italic since 100~keV electrons can travel further in this atmosphere than calculated within $\pm30\arcsec$.}
	\label{tb1}
	\end{center}
	\end{table*}
	\begin{figure*}[thpb]
	\centering
	\includegraphics[width=1.00\linewidth]{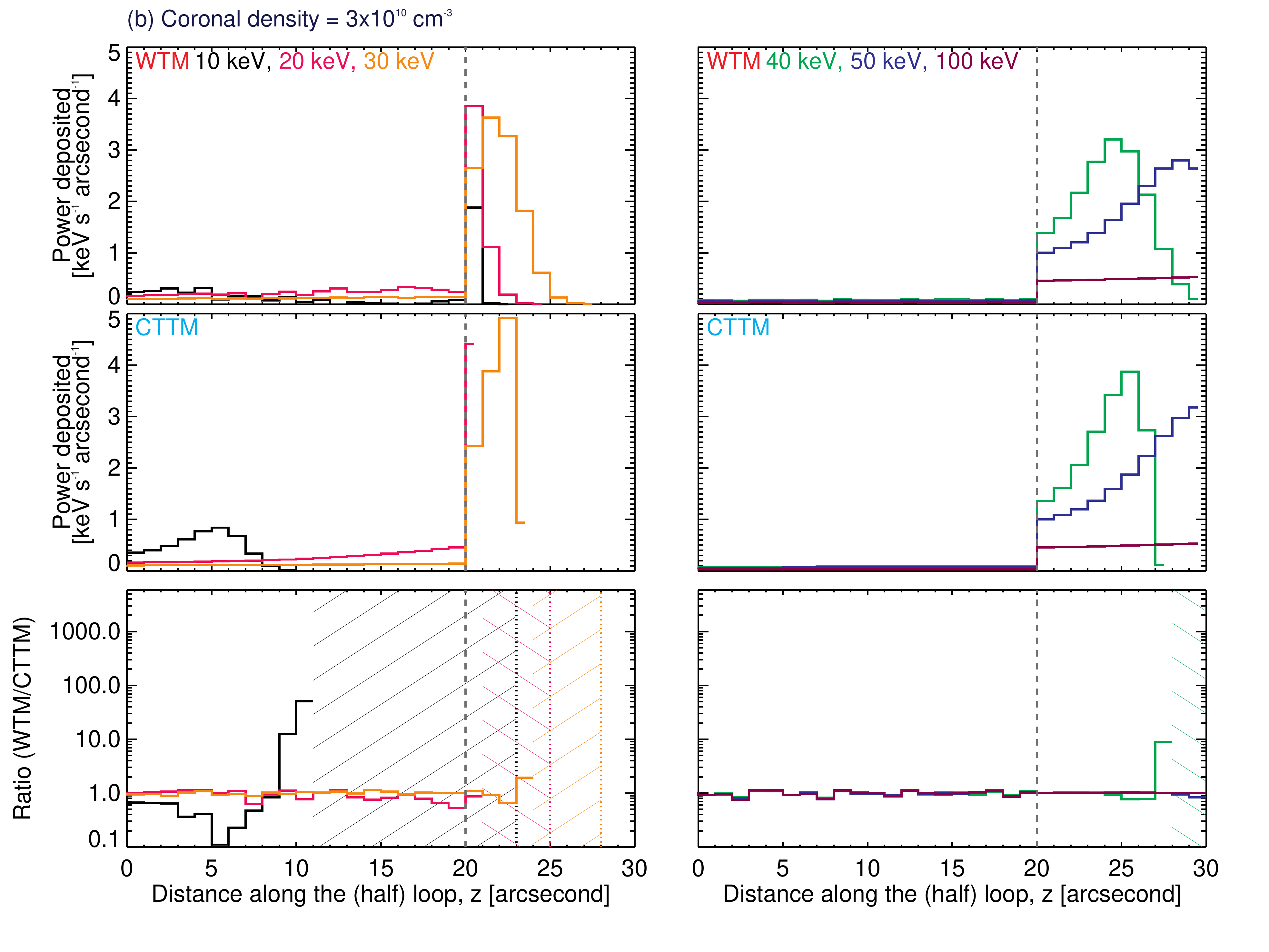}
	\caption{{\it Lower density flaring corona}. The results for (initally) mono-energetic simulation set (b) showing the spatial distribution of the electron power deposition in a WTM (top) and CTTM (middle), for different injected electron energies (10, 20, 30, 40, 50, 100 keV). The ratio of WTM result to CTTM result is shown in the bottom panel. The shaded regions in the bottom panels show regions where the power is deposited at greater depths in the WTM compared to the CTTM (and hence ratio$\rightarrow\infty$). Dashed grey line: corona-chromosphere boundary. Simulation set (b) uses a coronal temperature of $T=20$~MK, coronal loop length of $L=20\arcsec$, and a beamed injection.}
	\vspace{-10pt}
	\label{fig4}
	\end{figure*}
	\begin{table*}[thpb!]{}
	\begin{center}
	\footnotesize
	\begin{tabular}{|c|cc|cc|}
	\hline
	Set (b)  & \multicolumn{2}{c|}{CTTM} & \multicolumn{2}{c|}{WTM (T=20~MK)} \\ \hline
	\begin{tabular}{l}Energy {[keV] (E/T)}\end{tabular} & \begin{tabular}{l}Chromosphere ($\%$)\end{tabular} & \begin{tabular}{l}Corona ($\%$) \end{tabular} & \begin{tabular}{l}Chromosphere ($\%$) \end{tabular} & \begin{tabular}{l}Corona  ($\%$)\end{tabular} \\ \hline
	10 (5.8) & 0 & 100 & 46.8 & 53.2 \\
	20 (11.6) & 48.3 & 51.7 & 55.6 & 44.4 \\
	30 (17.4) & 83.8 & 16.2 & 83.6 & 16.4 \\
	40 (23.2) & 91.5 & 8.5 & 91.5 & 8.5 \\
	50 (29.0) & 93.5 & 6.5 & 93.5 & 6.5 \\
	100 (57.9) & \textit{88.7} & \textit{11.3} & \textit{88.7} & \textit{11.3}\\
	\hline
	\end{tabular}
	\caption{The percentage of available non-thermal electron power deposited in the corona and chromosphere for set (b) shown in Figure \ref{fig4}. The 100~keV values are shown in italic since 100~keV electrons can travel further in this atmosphere than calculated within $\pm30\arcsec$.}
	\label{tb2}
	\end{center}
	\end{table*}
	\begin{figure*}[thpb]
	\centering
	\includegraphics[width=0.49\linewidth]{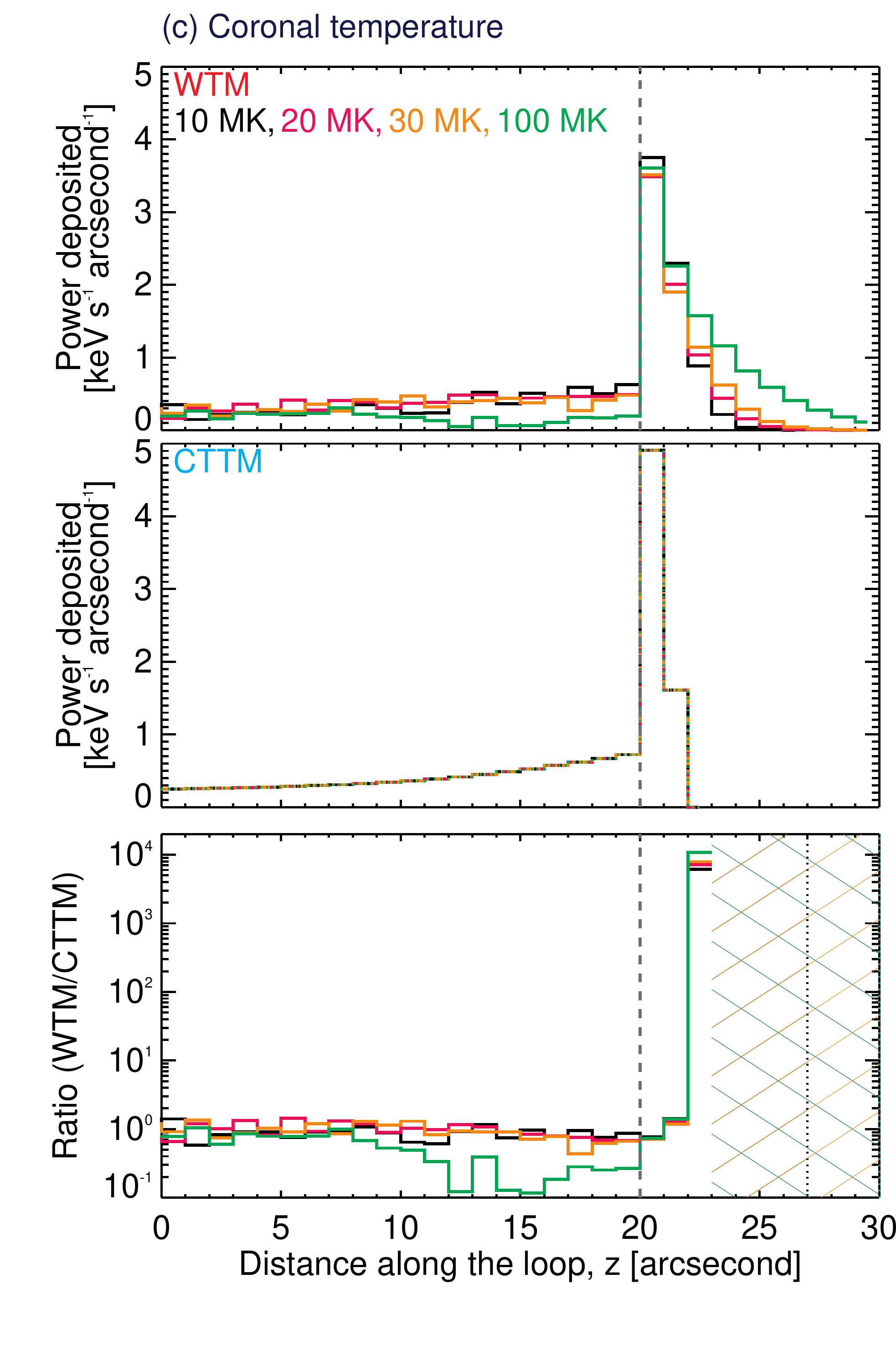}
	\includegraphics[width=0.49\linewidth]{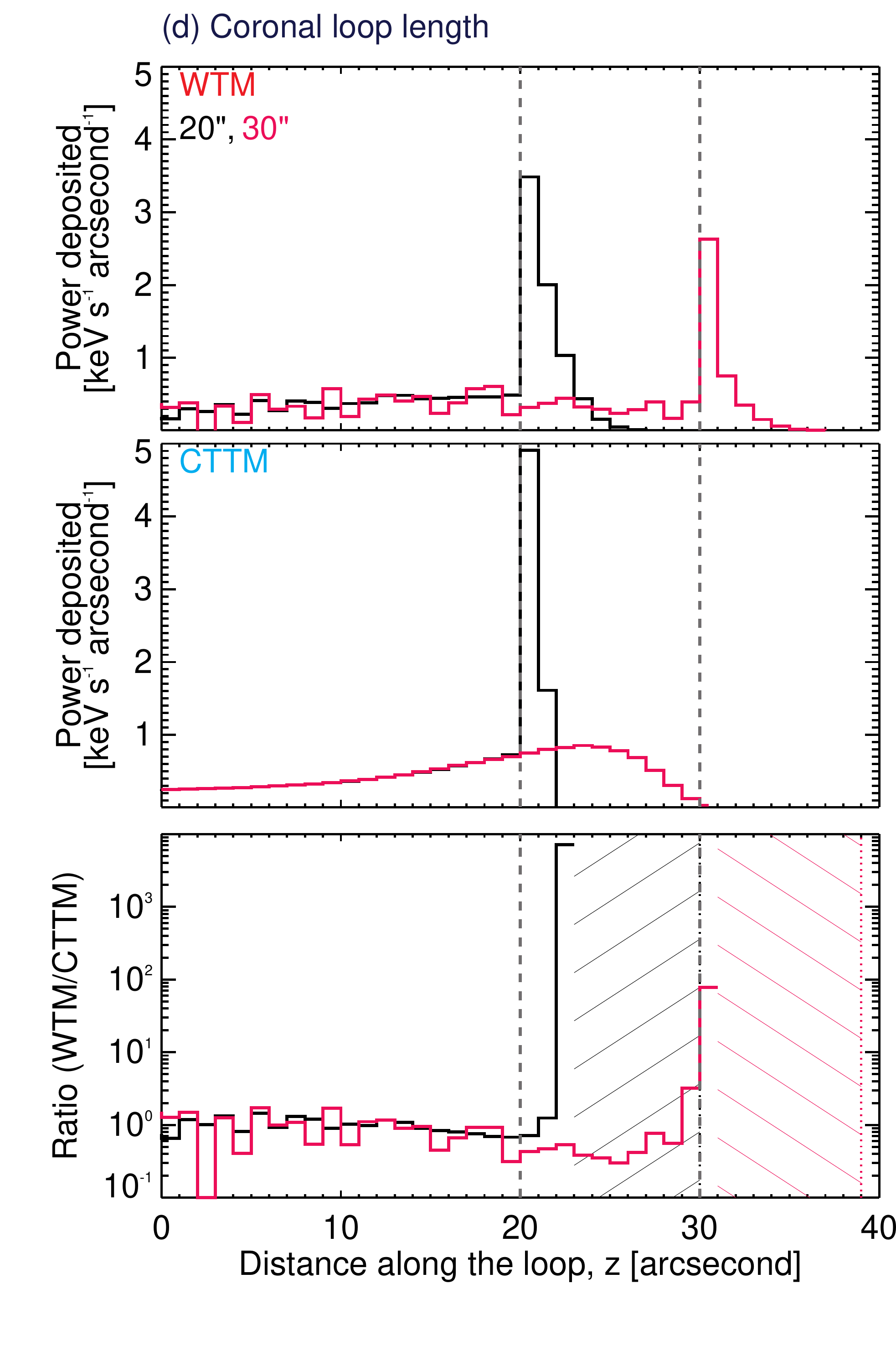}
	\caption{{\it Left - Different coronal temperatures, Right - Different coronal loop lengths}. The results for the (initially) mono-energetic (30~keV) sets (c; left) and (d; right) showing the spatial distribution of deposition in a WTM (top) and CTTM (middle). Dashed grey line: corona-chromosphere boundary. Simulation sets (c) and (d) use a coronal density of: $n=7\times10^{10}$~cm$^{-3}$, and a beamed injection. Set (c) uses a coronal loop length of $L=20\arcsec$ and set (d) uses a coronal temperature of $T=20$~MK. The ratio (WTM/CTTM) is shown in the bottom panel. The shaded regions in bottom panels show regions where the energy is deposited at greater depths in the WTM compared to the CTTM (and hence ratio$\rightarrow\infty$).}
	\vspace{-10pt}
	\label{fig5}
	\end{figure*}
	\begin{table*}[thpb]{}
	\begin{center}
	\footnotesize
	\begin{tabular}{|c|cc|cc|}
	\hline
 	Set (c) & \multicolumn{2}{c|}{CTTM} & \multicolumn{2}{c|}{WTM (T=20~MK)} \\ \hline
	\begin{tabular}{l}Temperature {[MK]}\;\;\; \\(E=30~keV/T)\end{tabular} & \begin{tabular}{l}Chromosphere ($\%$)\end{tabular} & \begin{tabular}{l}Corona ($\%$) \end{tabular} & \begin{tabular}{l}Chromosphere ($\%$) \end{tabular} & \begin{tabular}{l}Corona  ($\%$)\end{tabular} \\ \hline
	10 (34.8) & 47.1 & 52.9 & 51.8 & 48.2 \\
	20 (17.4) & 47.1 & 52.9 & 50.7 & 49.3 \\
	30 (11.6) & 47.1 & 52.9 & 53.5 & 46.5 \\
	100 (3.5) & 47.1 & 52.9 & 76.6 & 23.4 \\
	\hline
	\end{tabular}
	\bigskip

	\begin{tabular}{|c|cc|cc|}
	\hline
 	\noindent Set (d) & \multicolumn{2}{c|}{CTTM} & \multicolumn{2}{c|}{WTM (T=20~MK)} \\ \hline
	\begin{tabular}{l}Half loop length {[\arcsec]}\;\;  \\(E/T)\end{tabular} & \begin{tabular}{l}Chromosphere ($\%$)\end{tabular} & \begin{tabular}{l}Corona ($\%$) \end{tabular} & \begin{tabular}{l}Chromosphere ($\%$) \end{tabular} & \begin{tabular}{l}Corona  ($\%$)\end{tabular} \\ \hline
	20 (17.4) & 47.1 & 52.9 & 50.7 & 49.3 \\
	30 (17.4) & 0.7 & 99.3 & 28.3 & 71.7  \\
	\hline
	\end{tabular}
	\caption{The percentage of available non-thermal electron power deposited in the corona and chromosphere for set (c) and set (d) shown in Figure \ref{fig5}.}
	\end{center}
	\label{tb3}
	\end{table*}

\subsection{Mono-energetic energy inputs: temporal distribution of deposited power}\label{monot}

In all simulation sets, we also determine how long it takes for all 
of the non-thermal electron power to be deposited in the flaring atmosphere. Firstly, in Figure \ref{fig8}, we plot the results of sets (a) and (b). 
On each plot, we also add another simulation run where we inject (beamed) thermal electrons into the simulation. Note, that in the simulations, the injected electrons relax 
to the flux-averaged mean energy $2k_{B}T$ of the background plasma, 
and a $T=20$~MK corona gives $2k_{B}T=3.4$~keV. 
Figure \ref{fig7} shows that in a WTM, lower energy electrons $E/k_{B}T\le11.6$ can deposit their power in the lower atmosphere over a large range of timescales, and as expected the WTM result tends to the CTTM result when $E>50$ keV. 
These timescales are shown in Figure \ref{fig8} for both $n=1\times10^{11}$~cm$^{-3}$ and $n=3\times10^{10}$~cm$^{-3}$ cases. In the WTM, many electrons still deposit 
their power at times close to CTTM values, and WTM times converge to CTTM times as the electron energy increases. 
However, in the WTM, lower energy electrons show a large tail of delayed deposition in the lower atmosphere (of seconds to tens of second), unaccounted for in the CTTM, due to partially and fully thermalized electrons.

As an illustrative example from Figure \ref{fig8}, a 10~keV electron will thermalize quickly in a high density corona ($n=10^{11}$ cm$^{-3}$) over a distance of $<10\arcsec$ (over a time of $<0.1$~s). In the CTTM, 10~keV electrons in this scenario never reach the chromosphere and deposit energy. However, in a WTM, once a 10~keV electron has thermalized, it could travel hundreds of arcseconds 
in a random walk, in the hot coronal region, before exiting into the cooler and denser chromosphere. Therefore, depending on their path (the amount of scattering) 
and thermalization time, electrons can now exit the corona over a range of timescales from sub-second to tens of seconds\footnote{These times will be even further affected by turbulent scattering in the corona.}.

	\begin{figure*}[htpb]
	\centering
	\includegraphics[width=0.49\linewidth]{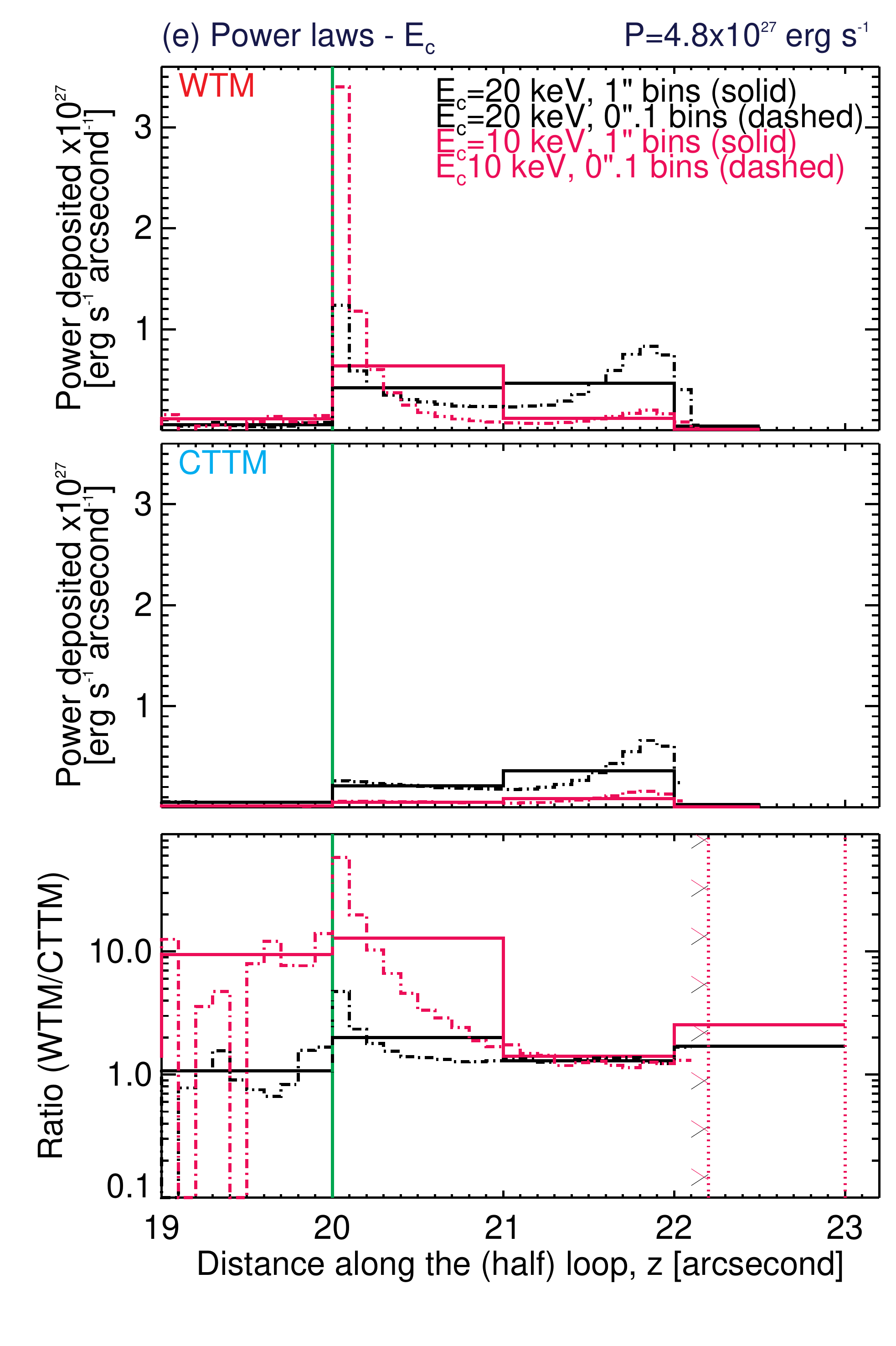}
	\includegraphics[width=0.49\linewidth]{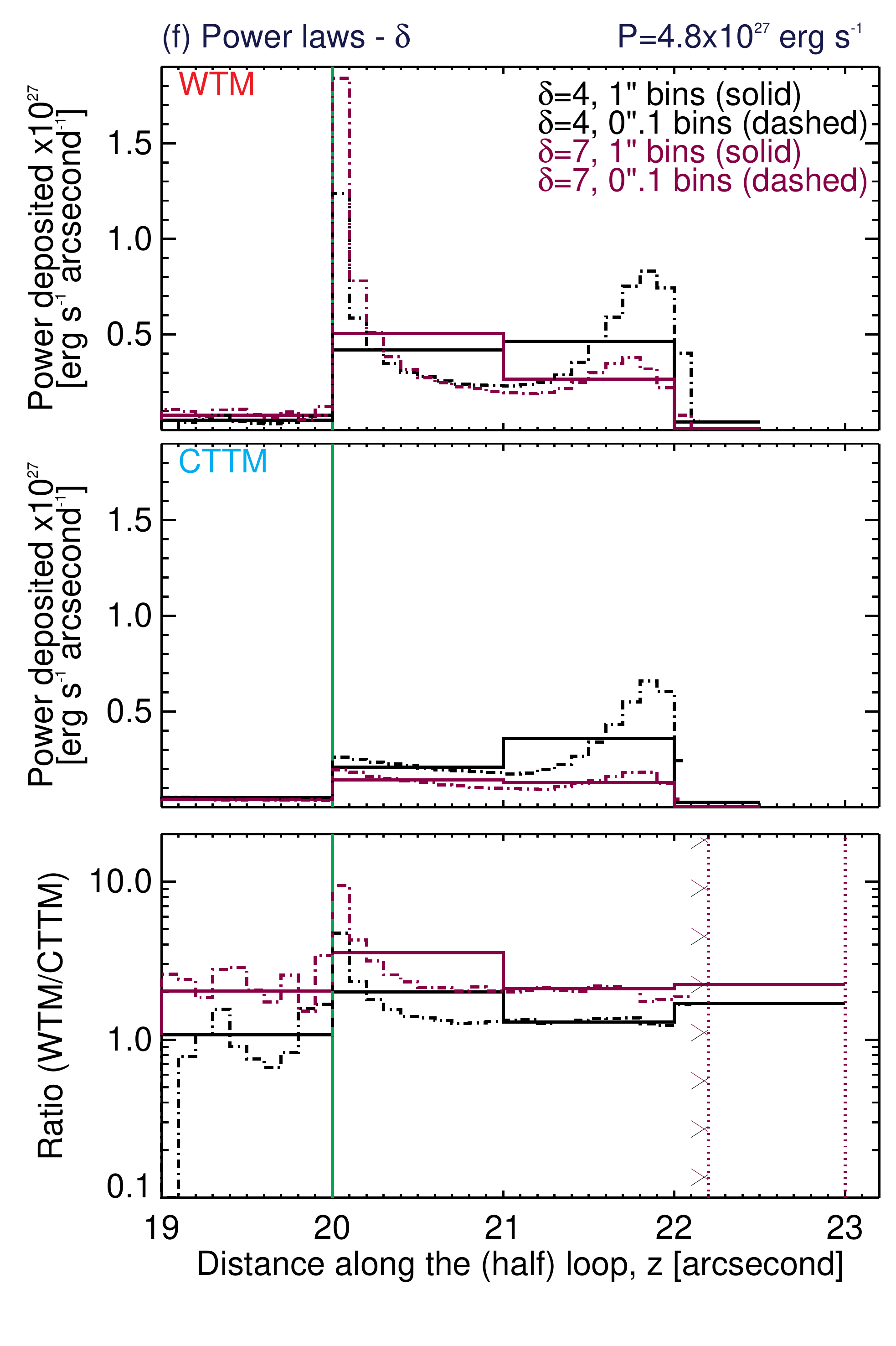}
	\vspace{-10pt}
	\caption{{\it Power law spectra; changes in electron power deposition with low energy cutoff and spectral index}. The results for sets (e; left) and (f; right) showing the spatial distribution of the electron power deposition (lower atmosphere only) in a WTM (top) and CTTM (middle). The ratio (WTM/CTTM) is shown in the bottom panel. Set (g) uses $E_{c}=10$~keV and $E_{c}=20$~keV and $\delta=4$. Set (h) uses $\delta=4$ and $\delta=7$, with energies ranging between $20-50$~keV.  All runs have the same injected non-thermal electron power of $P=4.8\times10^{27}$~erg/s. The shaded regions in bottom panels show regions where the energy is deposited at greater depths in the WTM compared to the CTTM (and hence ratio$\rightarrow\infty$). Green line: corona-chromosphere boundary. Simulation sets (e) and (f) use atmosphere type 2 and coronal parameters of: $T=20$~MK, $n=7\times10^{10}$~cm$^{-3}$ and $L=20\arcsec$.}
	\label{fig7}
	\end{figure*}
	\begin{table*}[htpb]{}
	\begin{center}
	\footnotesize
	\begin{tabular}{|c|cc|cc|}
	\hline
 	Set (e; 1\arcsec bins) & \multicolumn{2}{c|}{CTTM} & \multicolumn{2}{c|}{WTM (T=20~MK)} \\ \hline
	\begin{tabular}{l}$E_{c}$ {[keV] (E$_{c}$/T)}\;\;\;\end{tabular} & \begin{tabular}{l}Chromosphere ($\%$)\end{tabular} & \begin{tabular}{l}Corona ($\%$) \end{tabular} & \begin{tabular}{l}Chromosphere ($\%$) \end{tabular} & \begin{tabular}{l}Corona  ($\%$)\end{tabular} \\ \hline
	20 (11.6) & 29.5 & 70.5 & 47.5 & 52.6 \\
	10 (5.8) & 6.5 & 93.5 & 36.3 & 63.7   \\
	\hline
	\end{tabular}
	\bigskip

	\begin{tabular}{|c|cc|cc|}
	\hline
	\noindent Set (f; 1\arcsec bins) & \multicolumn{2}{c|}{CTTM} & \multicolumn{2}{c|}{WTM (T=20~MK)} \\ \hline
	\begin{tabular}{l}$\delta$\; {[keV] (E$_{c}$/T)}\;\;\;\;\end{tabular} & \begin{tabular}{l}Chromosphere ($\%$)\end{tabular} & \begin{tabular}{l}Corona ($\%$) \end{tabular} & \begin{tabular}{l}Chromosphere ($\%$) \end{tabular} & \begin{tabular}{l}Corona  ($\%$)\end{tabular} \\ \hline
	4 (11.6) & 29.5 & 70.5 & 47.5 & 52.6 \\
	7 (11.6) & 12.7 & 87.3 & 35.0 & 65.0 \\
	\hline
	\end{tabular}
	\caption{The percentage of available non-thermal electron power deposited in the corona and chromosphere for set (e) and set (f) shown in Figure \ref{fig7}.}
	\label{tb5}
	\end{center}
	\end{table*}

	\begin{figure*}[t]
	\centering
	\includegraphics[width=0.39\linewidth]{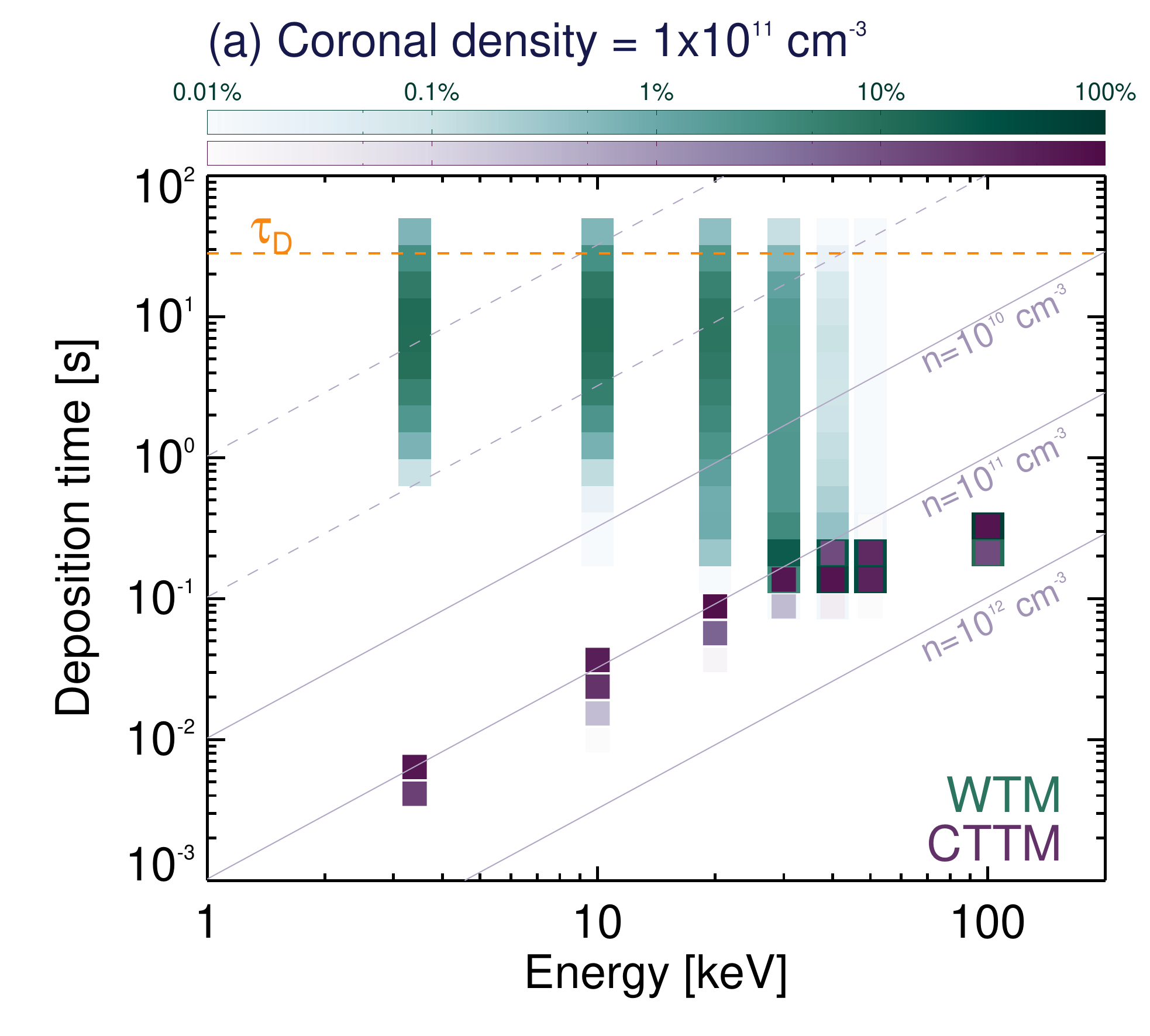}
	\includegraphics[width=0.39\linewidth]{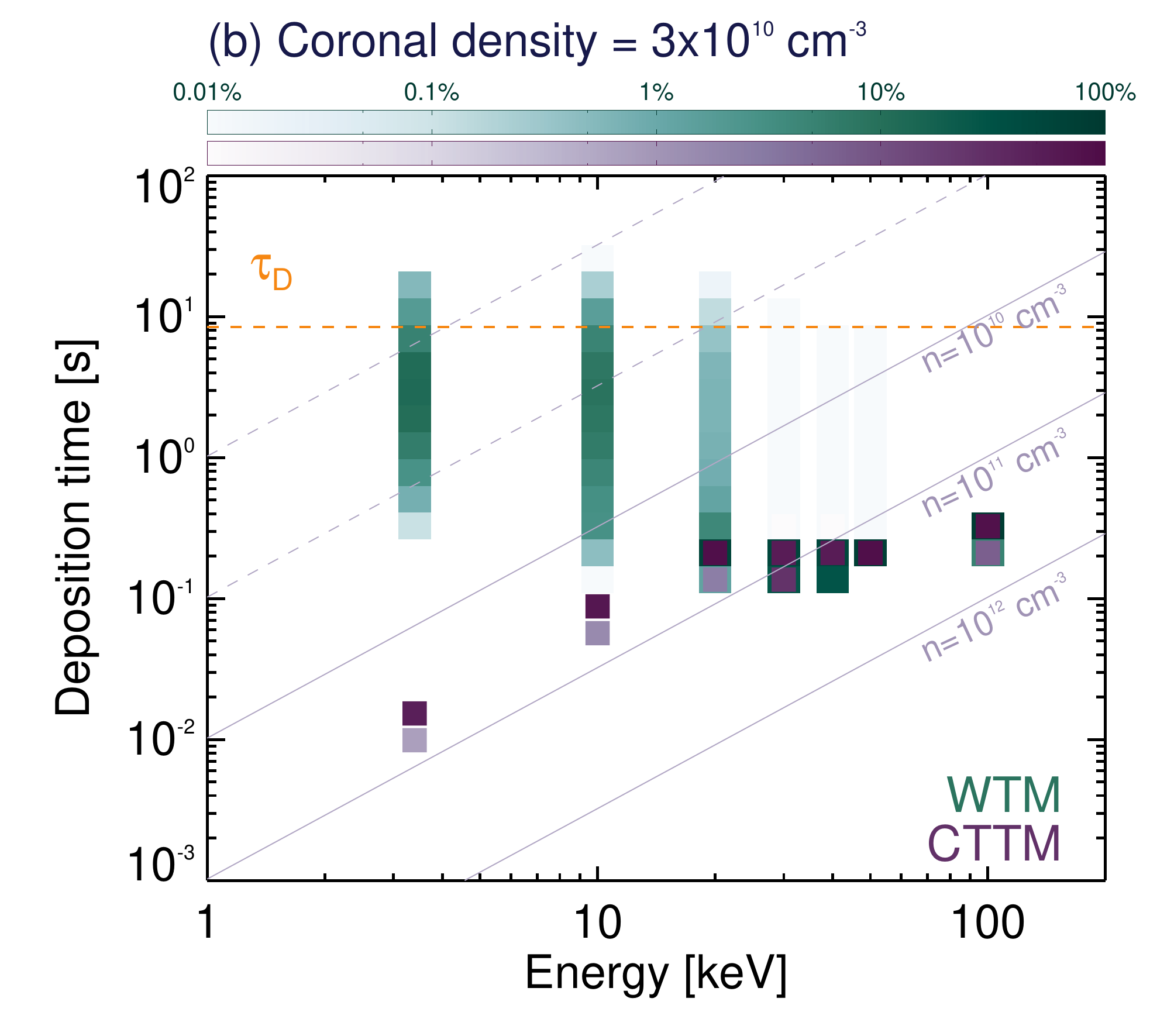}
	\includegraphics[width=0.39\linewidth]{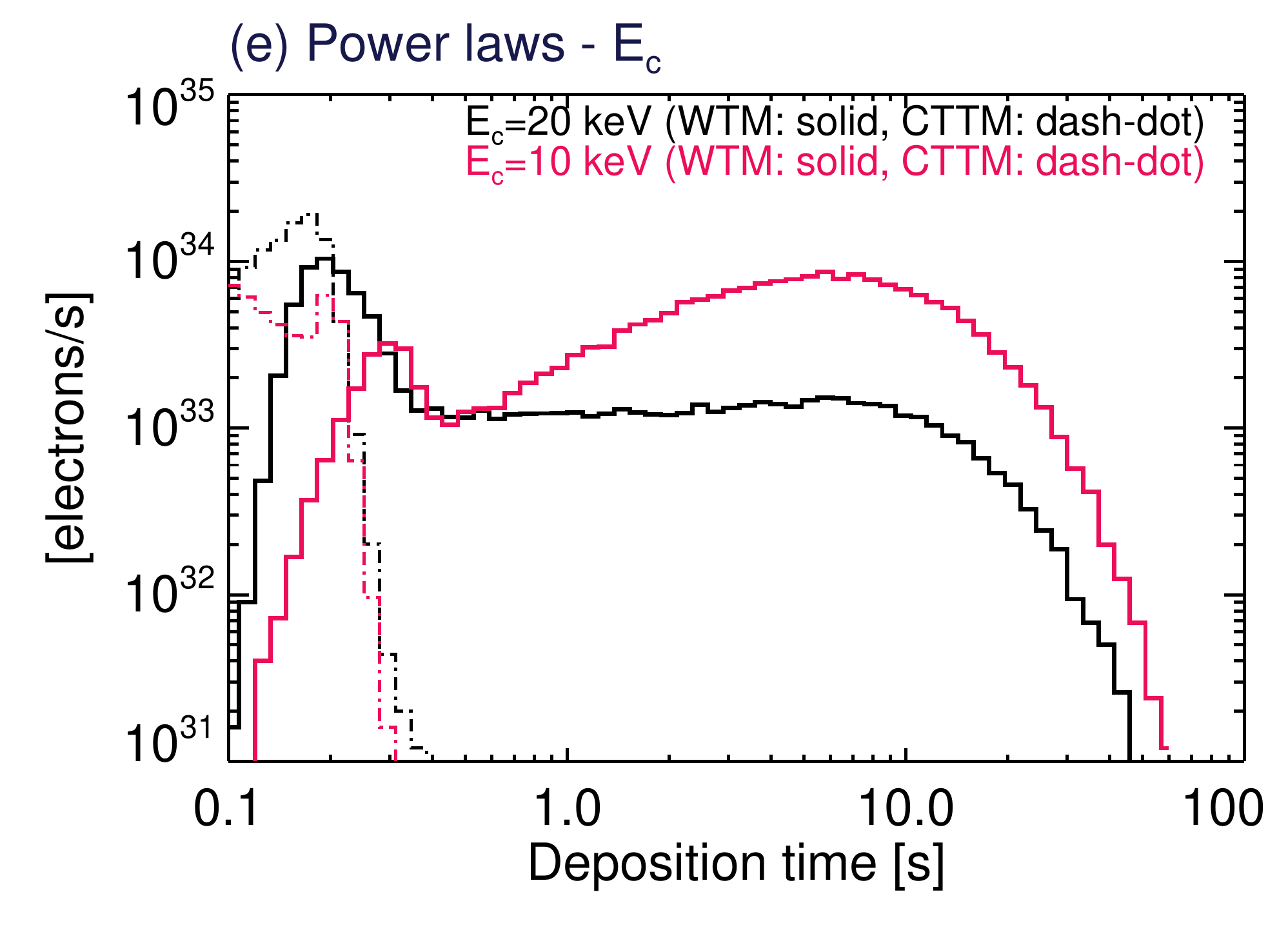}
	\includegraphics[width=0.39\linewidth]{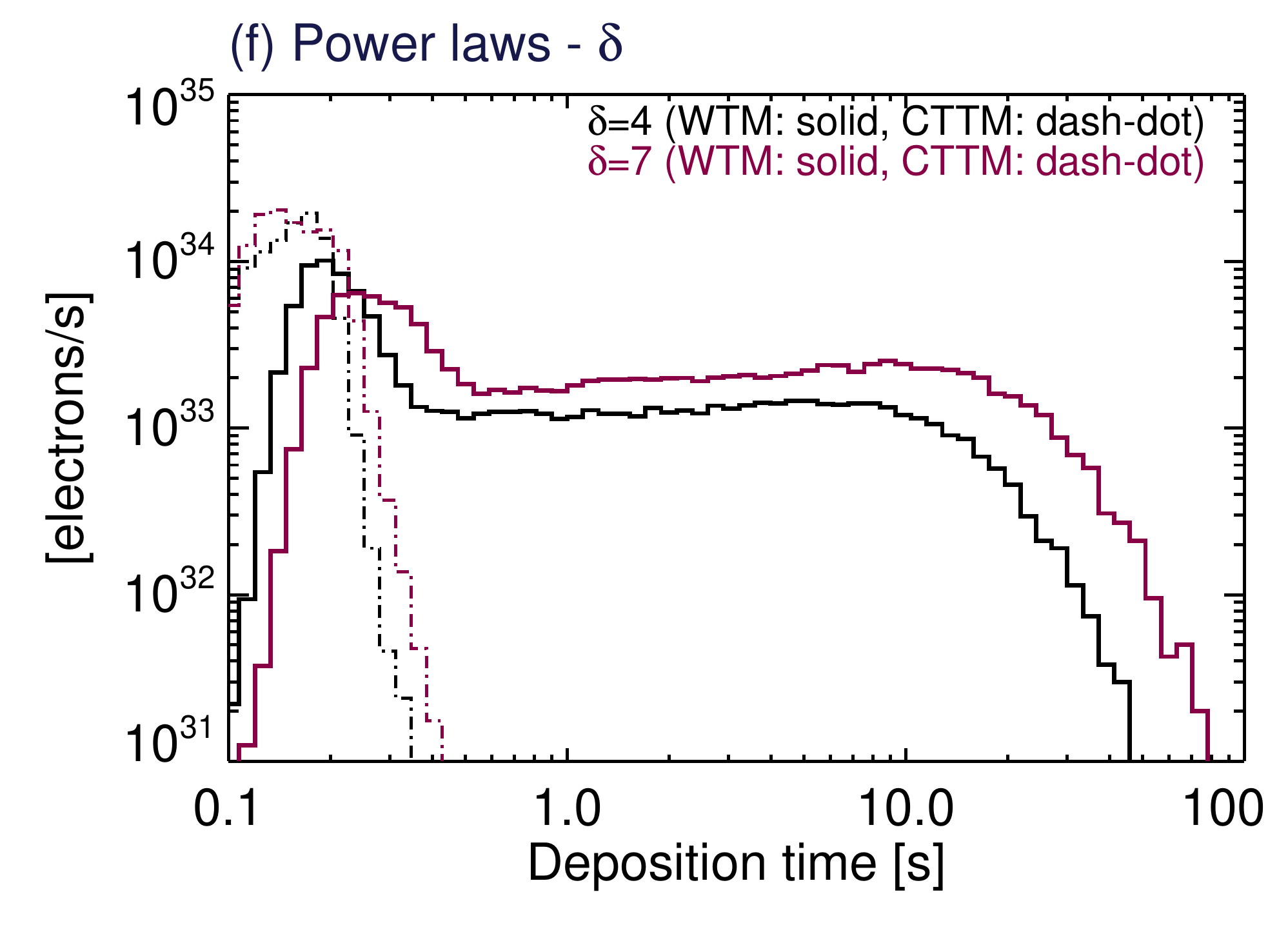}
	\caption{\textit{Top row:} Electron deposition times for simulation sets (a) and (b) (with the inclusion of thermal $2k_{B}T=3.4$ keV electrons); color bar indicates (e/s), for electrons in a WTM (green) and CTTM (purple). In a WTM, low energy electrons that thermalize in the corona reach and deposit their energy in the chromosphere over longer timescales of second to tens of seconds. Orange line: $\tau_{D}=$ thermal diffusion time. Purple lines: analytic energy loss times in a CTTM for a given density. \textit{Middle and bottom rows:} Electron deposition times for selected simulation sets (e) and (f). The shape of the delayed deposition curves is dependent on both the properties of the non-thermal electrons and coronal plasma properties.}
	\label{fig8}
	\end{figure*}
\subsection{Isotropic injection}
For cases where we input an isotropic electron distribution, we find similar results: more energy is deposited in the lower atmosphere in the WTM than in a CTTM. For example, in one simulation where we inject 30~keV electrons into a corona ($T=20$~MK, $n=7\times10^{10}$~[cm$^{-3}$] and $L=20\arcsec$), independent of whether the injected electron distribution is beamed or isotropic, more energy is deposited at deeper locations in the lower atmosphere in a WTM than in a CTTM, up to 1000 times more in the beamed case, at certain locations, and up to 4 times in the isotropic case. Moreover, initially isotropic non-thermal distributions deposit energy over a greater range of timescales, and more energy is deposited at greater times relative to a beamed injection of electrons, possibly providing a diagnostic of electron anisotropy.

\subsection{Solar flare power-law energy inputs}\label{presults}

To investigate how the power of flare-accelerated electrons is transferred 
and deposited in a more realistic solar or stellar flare scenario, 
we also perform simulation runs using an injected electron power-law energy distribution of the form $F(E_{0})\sim E_{0}^{-\delta}$ (set (e): different $E_{c}$ and set (f): different $\delta$). 
In these runs, we use the following plasma parameters and injected electron inputs: $T=20$~MK, $n=7\times10^{10}$~cm$^{-3}$ and the more realistic 
atmosphere type 2. 
Each electron distribution has a total power of $P=4.8\times10^{27}$~erg s$^{-1}$, with spectral index $\delta=4$ or $\delta=7$, 
a low energy cutoff of $E_{c}=10$~keV or $E_{c}=20$~keV and a high energy cutoff of $E_{H}=50$~keV. Again, such a low high-energy cutoff is used since we want to examine low-energy electrons that are incorrectly modelled by the CTTM. The results are shown in Figure \ref{fig7} and Table \ref{tb5}.

For the power law energy inputs we find the following notable results:
\begin{enumerate}
\item As expected, the differences in CTTM and WTM power deposition follow the results of Section \ref{monos} and Section \ref{monot}.
\item For a given injected electron power, we see a larger difference in CTTM and WTM deposition for electron distributions with a smaller low-energy cutoff $E_{c}$. Spatially, more power is deposited at greater depths and up to 12 times more power can be deposited in the lower atmosphere for $E_{c}=10$~keV and 2 times more energy for $E_{c}=20$~keV, at a given location, for the studied conditions.
\item For a given injected electron power, we see a larger difference in CTTM and WTM deposition for electron distributions with softer spectral indices. Spatially more power is deposited at greater depths and up to 4 times more power can be deposited 
    in the lower atmosphere for $\delta=7$ and up to 2 times for $\delta=4$, at a given location, for the studied conditions.
\item The simulation runs using $0\arcsec.1$ binning show that thermalized 
low energy electrons deposit a large fraction of power at the top of the chromospheric boundary.
\item Electron power law distributions with a higher fraction of low energy electrons (i.e smaller $E_{c}$ or larger $\delta$), 
    that partially or fully thermalize in the corona, deposit more of their energy at greater times (second to tens of second timescales).
\end{enumerate}

\section{Discussion}\label{discuss}

In this work, we show that the cold thick-target model (CTTM) does not adequately approximate the transport and deposition of energetic electrons in flaring, and therefore strongly heated, solar 
or stellar atmospheres. In the CTTM, neglecting second order effects 
such as velocity diffusion leads to an underestimate of the energy transferred to the low atmosphere in the majority of cases, 
especially by lower energy electrons with $E<50$~keV (or equivalently $E/k_{B}T\le20$) that can fully or partially thermalize in the flaring corona and transfer energy diffusively. 
This leads to a difference in the spatial distribution of deposited non-thermal 
electron power in both the corona and cool layers of the low atmosphere. Understanding energy transfer by low energy electrons is important. 
Most solar flare non-thermal electron distributions are consistent with steeply decreasing power laws with the bulk of the power held by electrons with $E<50$~keV. Further, the thermalization of non-thermal electrons in the corona leads to the non-thermal electron power being deposited over a large range of times from sub-second to tens of second, producing delayed heating 
in the lower atmosphere. The temporal distribution of this heating profile 
could act as a diagnostic of both non-thermal electron properties 
(even electron anisotropy, since isotropic electrons will spend longer in the coronal plasma, leading to greater thermalization), and the plasma conditions in the corona 
(temperature, number density and the extent of hot coronal plasma), if it can be extracted from observation.
Using a full WTM description of energy transfer and deposition may be especially important for the analysis of stellar flares with higher coronal temperatures and densities. Also, the WTM description of energy deposition may be important for the study of microflares with low-energy $\le$10~keV accelerated electrons \citep[e.g., ][]{2017ApJ...844..132W} that can easily thermalize in the coronal plasma, but still produce heating in the lower atmosphere.

The development of the warm-target model (WTM) shows the important role 
coronal plasma properties play in determining the acceleration, 
transport and now deposition of flare-accelerated non-thermal electron power. 
Future X-ray observatories must aim to better constrain the plasma properties 
for this purpose. These simulation results, although applicable 
to archived {\it RHESSI} data, anticipate the launch 
of direct imaging X-ray missions, that will be able to provide a more detailed 
picture of the solar flare environment (temperature, density, `hot' plasma extent) in different regions of the flare, using better spatial and temporal resolution. 
Proposed missions such the \textit{Focussing Optics X-ray Solar Imager (FOXSI}\,; \citet{2017arXiv170100792C}) will have a high dynamic range 
and greater imaging spectroscopy capabilities. 
The delayed energy transfer by thermalized electrons should be observable 
by imaging low energy soft X-rays ($<5$~keV). Current high spectral, 
spatial and temporal observations with {\em Interface Region Imaging Spectrograph} ({\em IRIS\,}; \citet{2014SoPh..289.2733D}) in the transition 
region and chromosphere can be used to study the effects of delayed 
heating by partially or fully thermalized non-thermal electron distributions.

\citet{2015A&A...584A..89J} show that the flaring corona is made up of multiple loops of varying temperature and density. These varying plasma parameters have a huge effect on both the injected electron parameters and on the resulting energy deposition. Such variation must be taken into account in future modelling. Further, we must study a changing, dynamic atmosphere, 
since deposition by thermalized non-thermal electrons is modulated by changes in the coronal plasma properties. 
Importantly, we suggest that hydrodynamic models that use the CTTM approximation as an input should be re-evaluated, 
and eventually the WTM should replace any CTTM approximations. 
WTM energy deposition with the inclusion of extended loop turbulence 
and magnetic trapping will also be the subject of upcoming work. 
As stated, the appeal of the CTTM is its simple analytic form 
and the next step is to produce a semi-analytic WTM function that can be used by both the solar and stellar communities to the determine the deposition of energy in flaring atmospheres.

\acknowledgments
NLSJ, EPK \& LF gratefully acknowledge the financial support from the Science and Technology Facilities Council (STFC) Consolidated Grant ST/P000533/1.
The work is supported by an international team grant (``Solar flare acceleration signatures and their connection to solar energetic particles'' \url{http://www.issibern.ch/teams/solflareconnectsolenerg/}) from the International Space Sciences Institute (ISSI) Bern, Switzerland.


\end{document}